\documentclass[12pt]{article}
\usepackage[english]{babel}    
\usepackage[ansinew]{inputenc} 
\usepackage[T1]{fontenc}       
\usepackage{lmodern,microtype} 
\usepackage{amsmath,amsthm,amssymb,amsfonts} 
\usepackage{dsfont,mathrsfs,ushort} 
\usepackage{titlesec,titling}  
\usepackage[nohead]{geometry}  
\usepackage{setspace,caption}  
\usepackage{enumitem,booktabs} 
\usepackage[comma]{natbib} 
\usepackage{pstricks,pst-all}   
\usepackage{tikz}
\usepackage{pst-plot,pst-node,pst-3dplot}
\usepackage{hyperref,breakurl}
\usepackage{pgfplots}
\usepackage{graphicx}
\usepackage{caption}
\usepackage{subcaption}
\usepackage{longtable}
\usepackage[flushleft]{threeparttable}
\usepackage{algorithmic}
\setenumerate{label=\small(\roman*)}
\hypersetup{colorlinks=true,pdfnewwindow=true,pdfstartview=FitH,%
	pdftitle="Peer Effects",pdfauthor="Nail Kashaev and Natalia Lazzati",%
	urlcolor=blue!90!red!45!black,citecolor=blue!90!red!45!black,linkcolor=red!90!black}
\DeclareCaptionFont{fancy}{\bfseries\sffamily}
\allowdisplaybreaks

\providecommand{\psreset}{\psset{%
		linewidth=0.3pt,linestyle=solid,linecolor=black,
		dotsize=2.5pt,dotsep=2.5pt,arrowsize=4pt,
		fillstyle=none,fillcolor=white,
		showpoints=false,arrows=-,linearc=0,framearc=0,
		hatchsep=2pt,hatchwidth=0.2pt,nodesep=4pt,opacity=1}
	\psset{gridcolor=black!60, subgridcolor=black!30}
}

\psreset

\usepackage{graphicx}
\graphicspath{ {figures/} }

\titleformat{\section}[block]{\centering\large\bfseries\sffamily}{\thesection.}{0.5em}{}
\titleformat{\subsection}[block]{\flushleft\bfseries}{\thesubsection.}{0.5em}{}
\titleformat{\subsection}[block]{\flushleft\bfseries\sffamily}{\thesubsection.}{0.5em}{}
\titleformat{\subsubsection}[runin]{\normalsize\itshape}{\bfseries\upshape\sffamily\thesubsubsection.}{0.5em}{}[.--\:]
\renewcommand{\thesubsubsection}{\arabic{section}.\arabic{subsection}.\alph{subsubsection}}
\titlespacing{\section}{0ex}{10ex}{5ex}
\titlespacing{\subsection}{0in}{6ex}{3ex}
\titlespacing{\subsubsection}{0mm}{2ex}{0.5em}
\providecommand{\abstitle}[1]{{\par\vspace*{2ex}\small\bfseries\sffamily #1}\hspace*{1ex}}
\renewenvironment{abstract}%
{\begin{center}\begin{minipage}{0.8\linewidth}%
			\setlength{\parindent}{0.0em}\abstitle{Abstract}\small}%
		{\end{minipage}\end{center}\vfill\clearpage}

\newtheorem{proposition}{Proposition}[section]

\providecommand{\tr}{^{\prime}}

  \theoremstyle{remark}
  
  \theoremstyle{plain}
  
  \theoremstyle{definition}
  
\theoremstyle{plain}

  \theoremstyle{plain}
  
 \theoremstyle{definition}
  
  \theoremstyle{plain}

  \providecommand{\assumptionname}{Assumption}

  \providecommand{\definitionname}{Definition}
  \providecommand{\lemmaname}{Lemma}
  
  \providecommand{\remarkname}{Remark}
\providecommand{\corollaryname}{Corollary}
\providecommand{\theoremname}{Theorem}
\providecommand{\examplename}{Example}

\usepackage{xcolor}

\makeatother
\begin{document}
\title{Peer Effects in Random Consideration Sets\thanks{For their useful comments and suggestions we would like to thank Victor Aguiar, Roy Allen, Tim Conley, Paola Manzini, Marco Mariotti, and Salvador Navarro. We would also like to thank the audience at Workshop in Experiments, Revealed Preferences, and Decisions, the Inter University Colloquium on Analytical Research in Economics, the CARE-Colloquium series at University of Kansas, NAWM ES 2021, and the seminar presentation at El Colegio de Mexico.}}
\author{ 
	Nail Kashaev\\
	Western University\\
	nkashaev@uwo.ca
	\and
	Natalia Lazzati\\
	UC Santa Cruz\\
	nlazzati@ucsc.edu
	}
\date{This version: May 16, 2021\\
      First version: April 13, 2019, arXiv:1904.06742}
\maketitle

\begin{abstract}
\noindent We develop a dynamic model of discrete choice that incorporates peer effects into random consideration sets. We characterize the equilibrium behavior and study the empirical content of the model. In our setup, changes in the choices of friends affect the distribution of the consideration sets. We exploit this variation to recover the ranking of preferences, attention mechanisms, and network connections. These nonparametric identification results allow unrestricted heterogeneity across people and do not rely on the variation of either covariates or the set of available options. Our methodology leads to a maximum-likelihood estimator that performs well in simulations. We apply our results to an experimental dataset that has been designed to study the visual focus of attention. 
\bigskip \bigskip


\end{abstract}

\section{Introduction}

\noindent In the last few years, the basic rational choice model of decision-making has been revised in different ways, partly recognizing that cognitive factors might play an important role in determining the choices of people. These revisions have produced, among others, the consideration set models. In these models, people do not consider all the available options at the moment of choosing, but a subset of them. It is still an open question in the literature how the consideration sets are formed and what their main determinants are. We think that peer effects might be of first-order importance in the formation process of the consideration sets.\footnote{This possibility has been (explicitly or implicitly) discussed by other researchers in specific contexts ---e.g., the choices of peers may help us discover a new television show \citep{godes2004using}, a new welfare program \citep{caeyers2014peer}, a new retirement plan \citep{duflo2003role}, a new restaurant \citep{qiu2018learning}, or an opportunity to protest \citep{enikolopov2020social}.} To study this possibility, this paper builds a dynamic model where the choices of friends affect the subset of options that a person ends up considering. We show the empirical content of the model and apply our main results to an experimental dataset that has been designed to evaluate the visual focus of attention of people. In our application, peer effects have a large impact on the attention mechanism.

In the dynamic model we build, people are linked through a social network. At a randomly given time, a person gets the opportunity to select a new option out of a finite set of alternatives. The person sticks to her new option until the revision opportunity arises again. People do not consider all the available options at the moment of revising their selection. Instead, each person first forms a consideration set and then picks her most preferred option from it. The distinctive feature of the model is that the probability that a given alternative enters the consideration set depends on the number of friends that are currently adopting that option. This theoretical model leads to a sequence of choices that evolves through time according to a Markov random process. The dynamic system induced by our model has a unique equilibrium that specifies the fraction of time that each vector of joint choices prevails in the long run. 

In the consideration set models, the relative frequency by which a person chooses each alternative does not necessarily respect her ranking of preference. Formally, let us say that a person makes a mistake when, at the moment of choosing, she selects an alternative that is dominated by another option according to her preference order.  In our model, a person could seldom choose her most preferred option at equilibrium. The set of connections, or the network structure, shapes the nature and the strength of the mistakes that people make. Thus, our model can be thought of as a model of peer effects in mistakes. We show via example that having friends with similar preferences ---a phenomenon known as homophily--- reduces the frequency by which people make mistakes. In addition, the example shows that having more friends also results in fewer mistakes.  

Having established equilibrium existence and characterized equilibrium behavior, we consider a researcher who observes a long sequence of choices made by the members of the network. We show that all primitives of the model can be uniquely recovered. These primitives include the ranking of preferences, the attention mechanism (or consideration probabilities), and the network structure. There are two aspects of our nonparametric identification results that deserve special attention. First, we allow unrestricted heterogeneity across people with respect to all parts of the model. Second, in contrast to most other works on consideration sets, we do not rely on the variation of either covariates or the set of available options (or menu) to recover these components. 

In our dynamic model, the observed choices of network members are generated by a system of conditional choice probabilities: each of these conditional choice probabilities specifies the distribution of choices of a given person conditional on the choices of others (at the moment of revising her selection). The identification strategy we offer is a two-step procedure. First, we show how to identify the primitives of the model using these conditional choice probabilities. Second, we study identification of the conditional choice probabilities from observed data. 

To recover the primitives of the model from the conditional choice probabilities we exploit that, in our framework, changes in the choices of friends induce stochastic variation in the consideration set probabilities. We use this variation to recover the set of connections between the people in the network and their ranking of preferences. We then use this information to recover the attention mechanism of each person, i.e., the probability of including a specific option in the consideration set as a function of the number of friends who are currently choosing it. As we explained earlier, this identification result is nonparametric and allows for full heterogeneity across people regarding preferences and attention probabilities. 

To identify the conditional choice probabilities, we consider two datasets: continuous-time data and discrete-time data with arbitrary time intervals. These two datasets coincide in that they provide a long sequence of choices from people in the network. They differ in the timing at which the researcher observes these choices. In continuous-time datasets, the researcher observes people's choices in real time. We can think of this dataset as the ``ideal dataset.'' With the proliferation of online platforms and scanners, this sort of data might be available in some applications. In this dataset, the researcher directly recovers the conditional choice probabilities. In discrete-time datasets, the researcher observes the joint configuration of choices at fixed time intervals (e.g., the choice configuration is observed every Monday). In this case, the conditional choice probabilities are not directly observed or recovered, they need to be inferred from the data. Adding an extra mild condition, we show that the conditional choices are also uniquely identified. For this last result we invoke insights from \citet{blevins2017identifying,blevins2018identification}. Briefly, the reason for which the second dataset suffices to recover the primitives of the model lies on the fact that the transition rate matrix of a continuous-time process with independent revision times across people is rather parsimonious. In particular, the probability that two or more people revise their selected options at the same time is zero. As a consequence, the transition rate matrix has zeros in many known locations. The non-zero elements can then be nicely recovered and they constitute a one-to-one mapping with the conditional choice probabilities.

Our initial results rely on a few simplifying restrictions. In particular, the preferences of each person are deterministic; we assume that there exists one alternative (the default) that is picked if and only if nothing else is considered; and we let the distribution of consideration sets be multiplicatively separable across alternatives and the probability of including each option depends on the number (but not the identity) of the friends that selected that option. We extend our model in various directions to relax these assumptions. First, we consider random preferences (on top of random consideration sets). The network structure and the attention mechanism are identified without any new assumptions. The random preferences are identified if and only if each person has ``enough'' friends ---the actual number of friends needed depends on the number of options. Second, we analyze a model with no default. All the primitives of this model are identified if there are more than three options in the set of available alternatives. Finally, we consider a set of extensions where consideration sets are formed arbitrarily (e.g., no multiplicative separability assumption). In these extensions, different friends may have different effects on the attention mechanism of a given person. In this model, we still can uniquely recover preferences and the network structure. The consideration probabilities, in general, are partially identified. However, we show that different forms of symmetry between consideration set probabilities are sufficient for their identification.  

After presenting the main identification results, we propose a consistent maximum-likelihood estimator of all parameters of the model that uses discrete-time data. Our Monte Carlo simulations imply that our estimator performs well even for relatively small datasets. 

Last, but not least, we apply our model to the experimental dataset used by  \citet{bai2019predicting}, which has been designed to evaluate the visual focus of attention of people. In the experiment, a group of people seated around a circle were asked to play the Resistance game. This is a party game in which people communicate with each other to find the deceptive players. A tablet was placed in front of each player and it recorded the direction of the sight of the player every third of a second (i.e., we consider each player looking to her left, right, or center). The structural model we offer allows us to disentangle the effects of two key forces in the visual focus of attention of people: directional sight preferences and peer effects in gazing. Regarding directional sight preferences, our estimates provide very strong support to the left-to-right bias. That is, the observation that people tend to look to the left more often than to the right.\footnote{See, for example, the experimental findings of \citet{spalek2004supporting,spalek2005left} and \citet{reutskaja2011search}. The left-to-right bias has also been documented in marketing research \burl{https://www.nngroup.com/articles/horizontal-attention-original-research/}. See also \citet{maass2003directional} and reference therein.} (This observation has guided, among others, the design of many online platforms!) Regarding peer effects in gazing, our results support the observation that people redirect their visual attention by following the gaze orientation of others, a phenomenon called ``gaze following.''\footnote{See, for example, \citet{gallup2012visual}.} 

We finally relate our results with the existing literature. From a modeling perspective, our setup combines the dynamic model of social interactions of \citet{blume1993statistical,blume1995statistical} with the (single-agent) model of random consideration sets of \citet{manski1977structure} and \citet{manzini2014stochastic}. By adding peer effects in the consideration sets we can use variation in the choices of others as the main tool to recover preferences. The literature on identification of single-agent consideration set models has mainly relied on variation of the set of available options or menus. The latter includes \citet{aguiar2017random, ABDsatisficing16, brady2016menu, caplin2018rational, cattaneo2017random, horan2019random, lleras2017more, manzini2014stochastic}, and \citet{masatliogluRA}. (See \citealp{aguiar2019doesrandom} for a comparison of several consideration set models in an experimental setting.) Other papers have relied on the existence of exogenous covariates that shift preferences or consideration sets. The latter include \citet{barseghyan2019discrete, barseghyan2019heterogeneous, crawford2021survey, conlon2013demand, draganska2011choice, gaynor2016free, goeree2008limited, mehta2003price}, and \citet{roberts1991development}. Variation of exogenous covariates has also been used by \citet{abaluck2017consumers} via an approach that exploits symmetry breaks with respect to the full consideration set model. \citet{aguiar2020identification, crawford2021survey}, and \citet{dardanoni2020inferring} use repeated choices but do not allow for peer effects (i.e., they work with panel data).  

There is a vast econometric literature on identification of models of social interactions where choices of peers affect preferences but not the choice sets (see \citealp{blume2011identification, bramoulle2020peer, de2017econometrics}, and \citealp{graham2015methods} for comprehensive reviews of this literature). We depart from this literature in that in our framework the direct interdependence between choices (endogenous effects) is captured by consideration sets (choice sets), not preferences.\footnote{Implicitly, preferences in our model capture the so-called contextual effects -- effects of predetermined factors that are not affected by peers.} As we mentioned earlier, we can recover from the data the set of connections between the people in the network. In the context of linear models, a few recent papers have made progress in the same direction. Among them, \citet{blume2015linear, bonaldi2015empirical, de2018recovering}, and \citet{manresa2013estimating}. In the context of discrete-choice, \citet{chambers2019behavioral} also identifies the network structure but in their model peers do not affect consideration sets but directly change preferences (among other differences).

In the paper, we connect the equilibrium behavior of our model with the Gibbs equilibrium. This connection is similar to the one in \citet{blume2003equilibrium}. We also distinguish our approach with the more standard models of peer effects in preferences. To this end we embed the models of \citet{brock2001discrete, brock2002multinomial} into our dynamic revision process. 

Let us finally mention two other papers that incorporate peer effects in the formation of consideration sets. \citet{borah2018} do so in a static framework and rely on variation of menus for identification. \citet{lazzati2018diffusion} considers a dynamic model but the time is discrete and she focuses on two binary options that can be acquired together.

The rest of the paper is organized as follows. Section~\ref{M} presents the model, the main assumptions, and some key insights of our approach (including how it differs from the more standard approach of peer effects in preferences). Section~\ref{Equi} describes the equilibrium behavior. Section~\ref{MI} studies the empirical content of the model. Section~\ref{E} extends the initial idea to contemplate random preferences (besides random consideration sets), the case of no-default option, and more general formation process for the consideration sets. Section~\ref{sec: empitrical application} applies our model to experimental data on visual focus of attention. Section~\ref{C} concludes, and all the proofs are collected in Appendix~\ref{P}. Appendices~\ref{app: gibbs} and~\ref{app: simul} cover the Gibbs random field model and provide some simulation results, respectively. Appendix~\ref{app: robustness} contains additional results for our empirical application.

\section{The Model}\label{M}
\noindent This section describes the model and the main assumptions we invoke in the paper. It also compares our approach to the more standard model of peer effects in preferences.
\subsection{Social Network, Consideration Sets, and Choices}
\noindent \textbf{Network and Choice Configuration} There is a finite set of
people connected through a social network. The network is described by a
simple graph $\Gamma =\left( \mathcal{A},e\right)$, where $\mathcal{A}=\left\{ 1,2,\dots,A\right\}$ is the finite set of nodes (or people) and $e$ is the set of edges. Each edge identifies two connected people and the direction of the connection. For each Person $a\in \mathcal{A}$ her set of friends (or reference group) is defined as follows:
\begin{equation*}
\mathcal{N}_{a}=\left\{ a'\in \mathcal{A}:a'\neq a\text{
and there is an edge from }a\text{ to }a'\text{ in } e \right\}. 
\end{equation*}
There is a set of alternatives $\overline{\mathcal{Y}}=\mathcal{Y}\cup\left\{ o\right\}$, where $\mathcal{Y}=\left\{1,2,\dots,Y\right\}$ is a finite set of options and $o$ is a default option. Each Person $a$ has a strict preference order $\succ_{a}$ over the set of options $\mathcal{Y}$. All people agree in that the default option is the least preferred.\footnote{Although we assume the same default option for all people, all our results go through if people have different default options as far as the researcher knows the identity of the worst option for each person.}
We extend the analysis to random preferences in Section~\ref{RP} and fully relax the specification of the default option in Section~\ref{default}. We refer to a vector $\mathbf{y}=\left( y_{a}\right)_{a\in \mathcal{A}}\in \overline{\mathcal{Y}}^{A}$ as a choice configuration.\bigskip

\noindent \textbf{Choice Revision} We model the revision of choices as a standard continuous-time Markov process. In particular, we assume that people are endowed with independent Poisson alarm clocks with rates $\mathbf{\lambda }=\left( \lambda_{a}\right)_{a\in \mathcal{A}}$.\footnote{See \citet{blume1993statistical, blume1995statistical} for theoretical models that rely on Poisson alarm clocks and \citet{blevins2018identification} for a nice discussion of the advantages of this type of revision process from an applied perspective.} At randomly given moments (exponentially distributed with mean $1/\lambda_{a}$) the alarm of Person $a$ goes off.\footnote{That is, each Person $a$ is endowed with a collection of random variables $\left\{\tau_{n}^{a}\right\}_{n=1}^{\infty}$ such that each difference $\tau_{n}^{a}-\tau_{n-1}^{a}$ is exponentially distributed with mean $1/\lambda_{a}$. These differences are independent across people and time.} When this happens, the person selects the most preferred alternative among the ones she is actually considering. Formally, if $\mathcal{C}\subseteq\mathcal{Y}$ is her consideration set, then the choice of Person $a$ can be represented by an indicator function 
\begin{equation*}
\operatorname{R}_{a}\left( v \mid \mathcal{C}\right) =\mathds{1}\left( v\succ_{a}v'\text{ for all }v'\in \mathcal{C} \text{ and }v\in \mathcal{C}\right)
\end{equation*}
that takes value 1 if $v$ is the most preferred option in $\mathcal{C}$ according to $\succ_{a}$. If, at the moment of choosing, the consideration set of Person $a$ does not include any alternative in $\mathcal{Y}$, then the person selects the default option.\bigskip

\noindent \textbf{Peer Effects in the Formation of Consideration Sets} In our model, whether Person $a$ pays attention to a particular alternative depends on her own choice and the configuration of choices of her friends at the moment of revising her selection. We indicate by $\operatorname{Q}_{a}\left( v \mid \mathbf{y}\right)$ the probability that Person $a$ pays attention to alternative $v$ given a choice configuration $\mathbf{y}$. It follows that the probability of facing consideration set $\mathcal{C}$ has the form of
\begin{equation*}
\prod\nolimits_{v\in \mathcal{C}}\operatorname{Q}_{a}\left( v \mid \mathbf{y}\right) \prod\nolimits_{v\notin \mathcal{C}}\left( 1-\operatorname{Q}_{a}\left(
v \mid \mathbf{y}\right) \right).
\end{equation*}
By combining preferences and stochastic consideration sets, the probability that Person $a$ selects (at the moment of choosing) alternative $v\in\mathcal{Y}$ is given by 
\begin{equation}
\operatorname{P}_{a}\left( v \mid \mathbf{y}\right) =\operatorname{Q}_{a}\left( v \mid \mathbf{y}\right) \prod\nolimits_{v'\in \mathcal{Y}, v' \succ_{a} v}\left( 1-\operatorname{Q}_{a}\left( v' \mid \mathbf{y}\right) \right).  \label{a}
\end{equation}
The default option is assumed to be always considered. Since it is also assumed to be the worst for every person, the default option is chosen by the person if, and only if, nothing else is considered. Thus, the probability of selecting $o$ is just $\prod\nolimits_{v\in\mathcal{Y}}\left( 1-\operatorname{Q}_{a}\left( v \mid \mathbf{y}\right)
\right)$. Leaving aside peer effects, this process of formation of consideration sets is analogous to the one studied by \citet{manski1977structure} and \citet{manzini2014stochastic}. In Section~\ref{subsec: general consideration} we extend our analysis to a more general setting.\bigskip

\noindent Altogether, the three elements just described characterize our initial model of peer effects in random consideration sets. (As mentioned above, we consider later several extensions.) This model leads to a sequence of joint choices that evolves through time according to a Markov random process. An equilibrium in this model is an invariant distribution on the space of joint choices of the dynamic process, that describes the frequency by which each member of the group selects each option. We next present three key assumptions we invoke in the paper.  

\subsection{Main Assumptions} 
\noindent Our results build on three assumptions. Let $\operatorname{N}_{a}^{v}\left( \mathbf{y}\right)$ be the number of friends of Person $a$ who select option $v$ in choice configuration $\mathbf{y}$. Formally, 
\begin{equation*}
\operatorname{N}_{a}^{v}\left(\mathbf{y}\right) =\sum\nolimits_{a'\in 
\mathcal{N}_{a}}\mathds{1}\left(y_{a'}=v\right).
\end{equation*}
We indicate by $\left\vert \mathcal{N}_{a}\right\vert$ the cardinality of $\mathcal{N}_{a}$. The three assumptions are as follows.\bigskip

\noindent \textbf{(A1)} For each $a\in\mathcal{A}$, $v\in \mathcal{Y}$, and $\mathbf{y}\in \overline{\mathcal{Y}}^{A}$, $1>\operatorname{Q}_{a}\left( v \mid \mathbf{y}\right) >0$. 
\medskip

\noindent \textbf{(A2)} For each $a\in\mathcal{A}$, $\left\vert\mathcal{N}_{a}\right\vert>0$.
\medskip

\noindent \textbf{(A3)} For each $a\in\mathcal{A}$, $v\in\mathcal{Y}$, and $\mathbf{y}\in \overline{\mathcal{Y}}^{A}$, 
\begin{equation*}
\operatorname{Q}_{a}\left( v \mid \mathbf{y}\right) \equiv \operatorname{Q}
_{a}\left( v \mid y_{a},\operatorname{N}_{a}^{v}\left( \mathbf{y}\right)
\right) \text{ is strictly increasing in }\operatorname{N}_{a}^{v}\left( \mathbf{y}\right) .
\end{equation*}

Assumption A1 states that, for any choice configuration, the probability of considering each option is strictly positive and lower than one, independently on how many friends have selected that option. This assumption captures the idea that people can eventually pay attention to an alternative for various reasons that are outside the control of our model (e.g., watching an ad on television or receiving a coupon). It also allows people to eventually disregard any further consideration of a given option, including the one that she is currently adopting. Assumption A1 guarantees that each subset of options is (ex-ante) considered with nonzero probability. Assumption A2 requires each person to have at least one friend. Assumption A3 states that the probability that a given person pays attention to a specific option depends on the current choice of the person and the number (but not the identity) of friends that currently selected it. Assumption A3 also states that each person pays more attention to a particular option if more of her friends are adopting it.

\subsection{Insights of the Model}\label{Ins}
\noindent In our model, the probability that Person $a$ selects option $v$ (at the moment of choosing) given network configuration $\mathbf{y}$ takes the form of
\[
\operatorname{P}_{a}\left( v \mid \mathbf{y}\right) = \operatorname{Q}_{a}\left( v \mid y_{a},\operatorname{N}_{a}^{v}\left( \mathbf{y}\right)\right) 
\prod\nolimits_{v'\in \mathcal{Y}, v' \succ_{a} v}\left( 1-\operatorname{Q}_{a}\left( v' \mid y_{a},\operatorname{N}_{a}^{v'}\left( \mathbf{y}\right)\right)\right).
\]
Let us formally say a person makes a mistake whenever she selects a dominated option. Note that the preference order of Person $a$ over the available alternatives is fixed. The choices of friends simply redirect the attention of the person from one alternative to another one. As a consequence, at any given time, the probability that Person $a$ selects a low ranked alternative that is popular among her friends can be much higher than the probability of selecting her most preferred option. In summary, in our model, the network structure does not shape preferences (or payoffs across alternatives) but the strength of the mistakes that people make. We show later, via example, that having similar friends -- a phenomenon known as homophily -- can sharply reduce the frequency of mistakes that the person makes. Having more friends acts in the same direction. 

Our model markedly differs from the more standard model of peer effects in preferences. To illustrate some of the main differences, note that in the same dynamic revision process we can easily embed a model of peer effects in preferences by modifying the conditional choice probabilities. Let us assume that individual choices are path independent, in the sense that choosing from a large set of options can be broken up into choosing from smaller subsets (Luce's Choice Axiom). If we add peer effects in preferences into the resulting choice model, then the probability that option $v$ is chosen given network configuration $\mathbf{y}$ can be expressed as follows
\[
\operatorname{P}_{a}\left( v \mid \mathbf{y}\right) =\ \mathrm{U}_{a}^{v}\left(y_{a},\operatorname{N}_{a}^{v}\left(\mathbf{y}\right)\right)/(\sum\nolimits_{v'\in\overline{\mathcal{Y}}}\mathrm{U}_{a}^{v'}\left(y_{a},\operatorname{N}_{a}^{v'}\left(\mathbf{y}\right)\right)),
\]
where $\mathrm{U}_{a}^{v}\left(y_{a},\operatorname{N}_{a}^{v}\left(\mathbf{y}\right)\right)$ is the utility Person $a$ gets from alternative $v$ given her previous choice $y_a$ and the number of friends who also picked it, $\operatorname{N}_{a}^{v}\left(\mathbf{y}\right)$.\footnote{If we further assume, as in \citet{brock2001discrete, brock2002multinomial}, that the payoff Person $a$ gets from alternative $v$ is the sum of a deterministic private utility the person receives from the choice, $\delta_{a,v}$, a social component that depends on the number of friends that select that option, $\mathrm{S}_{a,v}\left(\mathbf{y}\right)\equiv \mathrm{S}_{a,v}\left(y_{a},\operatorname{N}_{a}^{v}\left( \mathbf{y}\right)\right)$, and an idiosyncratic term that is doubly exponentially distributed with index parameter $1$, then $\mathrm{U}_{a}^{v}\left(y_{a},\operatorname{N}_{a}^{v}\left(\mathbf{y}\right)\right)$ takes the familiar exponential form $\exp\left(\delta_{a,v}+\mathrm{S}_{a,v}\left(y_{a},\operatorname{N}_{a}^{v}\left(\mathbf{y}\right)\right)\right)$.} In this model, the choices of friends affect the value that Person $a$ assigns to each option. Thus, at each given time, the relative frequency by which Person $a$ selects option $v$ over option $v'$ reflects her relative expected utility of the first option with respect to the second one. It follows that the alternative with the highest expected utility is always selected more often. This is in sharp contrast with what happens in our peer effects model in consideration sets, as we explained above. 

Let us finally remark that, expressed as we did, these two models also have different empirical implications. In particular, in the random consideration set model, the probability that Person $a$ selects alternative $v$ does not vary if some of her friends switch from the default option to any option that is dominated by alternative $v$ according to $\succ_a$. In contrast, in the model of peer effects in preferences, the probability of selecting alternative $v$ strictly increases under the same scenario. It follows that if we can recover the conditional choice probabilities from the data (as we do in Section~\ref{MI}), then the two models can be set apart. 

\section{Equilibrium Behavior}\label{Equi}

\noindent The independent and identically distributed Poisson alarm clocks, which lead the selection revision process, guarantee that at each moment of time at most one person revises her selection almost surely. Thus, the transition rates between choice configurations that differ in more than one person changing the current selection are zero. The advantage of this fact for model identification is that there are fewer terms to recover. \citet{blevins2017identifying,blevins2018identification} offers a nice discuss of this feature and its advantage over discrete time models. Formally, the transition rate from choice configuration $\mathbf{y}$ to any different one $\mathbf{y}'$ is as follows
\begin{equation}
\operatorname{m}\left( \mathbf{y}' \mid \mathbf{y}\right) =\left\{ 
\begin{array}{lcc}
0 & \text{if} & \sum\nolimits_{a\in \mathcal{A}}\mathds{1}\left( y_{a}'\neq
y_{a}\right) >1 \\ 
\sum\nolimits_{a\in \mathcal{A}}\lambda_{a}\operatorname{P}_{a}\left( y_{a}' \mid \mathbf{y}\right) \mathds{1}\left( y_{a}'\neq y_{a}\right) & 
\text{if} & \sum\nolimits_{a\in \mathcal{A}}\mathds{1}\left( y_{a}'\neq
y_{a}\right) =1
\end{array}
\right. .  \label{T}
\end{equation}
In the statistical literature on continuous-time Markov processes these transition rates are the out of diagonal terms of the \textit{transition rate matrix} (also known as the \textit{infinitesimal generator matrix}). The diagonal terms are simply build from these other values as follows
\begin{equation*}
\operatorname{m}\left( \mathbf{y}\mid \mathbf{y}\right) =-\sum\nolimits_{\mathbf{y}%
'\in \overline{\mathcal{Y}}^{A}\setminus \left\{ \mathbf{y}\right\}
}\operatorname{m}\left( \mathbf{y}'\mid \mathbf{y}\right) .
\end{equation*}
We will indicate by $\mathcal{M}$ the transition rate matrix. In our model, the number of choice configurations is $\left(Y+1\right)^{A}$. Thus, $\mathcal{M}$ is a $\left(Y+1\right)^{A}\times \left(Y+1\right)^{A}$ matrix. There are many different ways of ordering the choice configurations and thereby writing the transition rate matrix. To avoid any sort of ambiguity in the exposition, we will let the choice configurations be ordered according to the lexicographic order with $o$ treated as zero. Constructed in this way the first element of $\mathcal{M}$ is $\mathcal{M}_{11}=\operatorname{m}\left(\left(o,o,\dots,o\right)\tr\mid \left(o,o,\dots,o\right)\tr\right)$. Formally, let $\iota\left( \mathbf{y}\right) \in\left\{1,2,\dots,\left(Y+1\right)^{A}\right\} $ be the position of $\mathbf{y}$ according to the lexicographic order. Then, 
\begin{equation*}
\mathcal{M}_{\iota \left(\mathbf{y}\right) \iota \left(\mathbf{y}'\right) }=\operatorname{m}\left(\mathbf{y}'\mid \mathbf{y}\right) .
\end{equation*}
\par
An equilibrium in this model is an invariant distribution  $\mu: \overline{\mathcal{Y}}^{A} \rightarrow\left[0,1\right]$, with $\sum\nolimits_{\mathbf{y}\in \overline{\mathcal{Y}}^{A}}\mu \left( \mathbf{y}\right) =1$, of the dynamic process with transition rate matrix $\mathcal{M}$. It indicates the likelihood of each choice configuration $\mathbf{y}$ in the long run. This equilibrium behavior relates to the transition rate matrix in a linear fashion 
\begin{equation*}
\mu \mathcal{M}=\mathbf{0}.
\end{equation*}
The next proposition establishes equilibrium existence and uniqueness for our model.
\begin{proposition}\label{P1}
If Assumption A1 is satisfied, then there exists a unique equilibrium $\mu$.
\end{proposition}

\noindent \textbf{Remark. }In Appendix~\ref{app: gibbs} we relate the equilibrium notion we use with the so-called Gibbs equilibrium. This alternative solution concept has been extensively used in Economics to study social interactions. (See \citealp{allen1982some, blume1993statistical, blume1995statistical}, and \citealp{blume2003equilibrium}, among many others.) The existence of Gibbs equilibrium requires extra symmetry restrictions among people. While these additional restrictions make the model more tractable, they are hard to justify in empirical applications. This is why we selected a more general approach.\bigskip

The first example describes the equilibrium behavior of a very simple specification of our model.\bigskip

\noindent \textbf{Example 1.} The network consists of two identical people that select between two alternatives, option $1$ and the default option $o$. The rates for their Poisson alarm clocks are $1$. Let us also assume, to simplify the setup, that the probability of paying attention to a particular option only depends on the current choice of the other person (i.e., $\operatorname{Q}_{a}\left( v \mid y_a, \operatorname{N}_{a}^{v}\left( \mathbf{y}\right) \right)=\operatorname{Q}_{a}\left( v \mid \operatorname{N}_{a}^{v}\left( \mathbf{y}\right) \right)$) and is the same for both people (i.e., $\operatorname{Q}(\cdot)=\operatorname{Q}_1(\cdot)=\operatorname{Q}_2(\cdot)$). Thus, for $a=1,2$, we get that 
\begin{equation*}
\operatorname{P}_{a}\left( 1 \mid \mathbf{y}\right) =\operatorname{Q}\left( 1 \mid \operatorname{N}_{a}^{v}\left( \mathbf{y}\right) \right) \text{ and }\operatorname{P}_{a}\left( o \mid \mathbf{y}\right) =1-\operatorname{Q}\left( 1 \mid \operatorname{N}_{a}^{v}\left( \mathbf{y}\right) \right).
\end{equation*}
The transition rate matrix $\mathcal{M}$ is as follows.
\begin{equation*}
\begin{tabular}{c|c|c|c|c|}
\multicolumn{1}{c}{}&\multicolumn{1}{c}{$\left(o,o\right)$}&\multicolumn{1}{c}{$\left(o,1\right)$}&\multicolumn{1}{c}{$\left(1,o\right)$}&\multicolumn{1}{c}{$\left(1,1\right)$}\\
\cline{2-5}
$\left( o,o\right)$ & $-2\operatorname{Q}\left( 1 \mid 0\right)$ & $\operatorname{Q}\left( 1 \mid 0\right)$ & $\operatorname{Q}\left( 1 \mid %
0\right)$ & $0$ \\ \cline{2-5}
$\left( o,1\right)$ & $1-\operatorname{Q}\left( 1 \mid 0\right)$ & $-1+%
\operatorname{Q}\left( 1 \mid 0\right) -\operatorname{Q}\left( 1 \mid %
1\right)$ & $0$ & $\operatorname{Q}\left( 1 \mid 1\right)$ \\ \cline{2-5}
$\left( 1,o\right)$ & $1-\operatorname{Q}\left( 1 \mid 0\right)$ & $0$ & $%
-1+\operatorname{Q}\left( 1 \mid 0\right) -\operatorname{Q}\left( 1 \mid %
1\right)$ & $\operatorname{Q}\left( 1 \mid 1\right)$ \\ \cline{2-5}
$\left( 1,1\right)$ & $0$ & $1-\operatorname{Q}\left( 1 \mid 1\right)$ & $%
1-\operatorname{Q}\left( 1 \mid 1\right)$ & $-2+2\operatorname{Q}\left( 1\mathcal{%
\ \mid }1\right)$ \\ \cline{2-5}
\end{tabular}%
\ .
\end{equation*}
The transition rate matrix $\mathcal{M}$ is naturally more complex when there are more alternatives or more people. However, the structure of the zeros in $\mathcal{M}$ is rather similar. In particular, there are many zeros in known locations. As we mentioned earlier, this feature of the model is particularly attractive for identification and estimation. 

The invariant distribution of choices, or steady-state equilibrium, satisfies 
$\mu\mathcal{M}=\mathbf{0}$.
Solving this system of equations, we get that the steady-state equilibrium is 
\begin{align*}
\mu \left( o,o\right) &=\frac{\left[ 1-\operatorname{Q}\left( 1 \mid 0\right) \right] \left[ 1-\operatorname{Q}\left( 1 \mid 1\right) \right] }{1-\operatorname{Q}\left( 1 \mid 1\right) +\operatorname{Q}\left( 1 \mid 0\right) }, \\
\mu \left( o,1\right) &=\mu \left( 1,o\right) = \frac{\operatorname{Q}\left( 1 \mid 0\right) \left[ 1-\operatorname{Q}\left( 1 \mid 1\right)\right] }{1-\operatorname{Q}\left( 1 \mid 1\right) +\operatorname{Q}\left( 1 \mid 0\right) }, \\
\mu\left( 1,1\right) &=\frac{\operatorname{Q}\left( 1 \mid 0\right) \operatorname{Q}\left( 1 \mid 1\right) }{1-\operatorname{Q}\left( 1 \mid 1\right) +\operatorname{Q}\left( 1 \mid 0\right) }.
\end{align*}
The steady-state equilibrium is a joint distribution on the pair of choice configurations. It states the fraction of time that each pair of choices $(y_{1},y_{2})$ prevails in the long run.{\normalsize\hfill }$\blacksquare $\bigskip

The second example shows how the strength of the mistakes that people make relates to the structure of the network. We will use the simulated data in the example later to show how the main parts of the model can, indeed, be estimated from a sequence of choices.
\bigskip

\noindent \textbf{Example 2.}\label{R} There are five people in the network. Their reference groups are as follows 
\[
\mathcal{N}_{1}=\left\{ 2\right\},\quad\mathcal{N}_{2}=\left\{ 1\right\},\quad\mathcal{N}_{3}=\left\{ 1,2\right\},\quad\mathcal{N}_{4}=\left\{ 5\right\} ,\quad\mathcal{N}_{5}=\left\{ 4\right\}.
\]
Note that each person has at least one friend, so Assumption A2 is satisfied. Moreover, the network is directed since $3\notin \mathcal{N}_{1},\mathcal{N}_{2}$ and $1,2\in\mathcal{N}_{3}$. There are two possible alternatives $\mathcal{Y} = \left\{ 1,2\right\}$ in addition to the default option $o$. The preferences of these people are as follows 
\[
2\succ_{1}1,\quad 1\succ_{2}2,\quad 2\succ_{3}1,\quad 1\succ_{4}2,\quad 1\succ_{5}2.
\]
That is, Persons 2, 4, and 5 prefer option 1, and Persons 1 and 3 prefer option 2. We will assume the attention mechanism is invariant across people and alternatives. To keep the exercise simple, as in Example 1, we will also assume the probability of paying attention to a given option only depends on the choices of friends. In this case, we indicate by $\operatorname{Q}\left( v \mid \operatorname{N}_{a}^{v}\left(\mathbf{y}\right) \right)$ the probability that Person $a$ pays attention to option $v\in \mathcal{Y}$ if the number of friends that are currently selecting that option is $\operatorname{N}_{a}^{v}\left( \mathbf{y}\right)$. We initially let 
\[
\operatorname{Q}\left( v \mid 0\right) =\frac{1}{4},\quad\operatorname{Q}\left( v \mid 1\right) =\frac{3}{4},\quad\operatorname{Q}\left( v \mid 2\right) =\frac{7}{8}.
\]
The rates for their Poisson alarm clocks are 1.

The equilibrium behavior of this choice model is a joint distribution $\mu $ with support on 243 choice configurations $\left(3^{5}\right)$. We simulated a long sequence of choices and calculated the equilibrium behavior. (See Appendix~\ref{app: simul} for more details.) From the equilibrium distribution we can obtain the marginal equilibrium probabilities for each person. Each of these marginals specifies the fraction of time that a given person selects each alternative in the long run. These marginal probabilities are shown in Table~\ref{table: marg prob 1}. 
\begin{table}
\centering
\begin{threeparttable}
\centering
\caption{Marginal Equilibrium Probabilities for Directed Network}\label{table: marg prob 1}
\begin{tabular}{|c|c|c|c|c|}
\hline
$\mu_{1}\left( o\right) =0.29$ & $\mu_{2}\left( o\right) =0.29$ & $\mu
_{3}\left( o\right) =0.19$ & $\mu_{4}\left( o\right) =0.30$ & $\mu
_{5}\left( o\right) =0.30$ \\ 
$\mu_{1}\left( 1\right) =0.30$ & $\mu_{2}\left( 1\right) =0.40$ & $\mu
_{3}\left( 1\right) =0.30$ & $\mu_{4}\left( 1\right) =0.50$ & $\mu
_{5}\left( 1\right) =0.50$ \\ 
$\mu_{1}\left( 2\right) =0.41$ & $\mu_{2}\left( 2\right) =0.31$ & $\mu
_{3}\left( 2\right) =0.51$ & $\mu_{4}\left( 2\right) =0.20$ & $\mu
_{5}\left( 2\right) =0.20$ \\ \hline
\end{tabular}%
\vspace{1ex}
\begin{tablenotes}
\item {\footnotesize Notes: These probabilities are estimated using simulated data, sample size $=15000$.}
\end{tablenotes}
\end{threeparttable}
\end{table}
It is interesting to observe that, for some people in the network, the relative frequency of choosing each alternative does not respect their order of preferences. In particular, notice that Persons 4 and 5 select the default more often than option 2, which dominates the former according to their preferences.

As before, let us say that a person makes a mistake when does not select her most preferred alternative. From the equilibrium marginals, we can calculate the probabilities of making mistakes. These probabilities are displayed in Table~\ref{table: prob of mistakes 1}.  
\begin{table}
\centering
\begin{threeparttable}
\centering
\caption{Probabilities of Mistakes for Directed Network}\label{table: prob of mistakes 1}
\begin{tabular}{|c|c|c|c|c|c|}
\hline
& Person 1 & Person 2 & Person 3 & Person 4 & Person 5 \\ \hline
Probability of Mistakes & $59$\% & $60$\% & $49$\% & $50$\% & $50$\% \\ \hline
\end{tabular}
\vspace{1ex}
\begin{tablenotes}
\item {\footnotesize Notes: These probabilities are computed using the estimated marginal equilibrium probabilities from Table~\ref{table: marg prob 1}.}
\end{tablenotes}
\end{threeparttable}
\end{table}
Note that Persons 2 and 4 are identical in all respects except in the type of friend they have. In particular, Person 4 shares with her friend the same preferences. The opposite is true for Person 2. This difference leads Person 4 to make fewer mistakes. Hence, having friends with similar preferences (homophily) seems to help each person to consider more often her best alternative. Also, note that Person 3 has more friends and makes fewer mistakes. Thus, having more friends is also helpful in this application. 

To show a bit more how the network structure shapes people's mistakes, we add two more connections in the model. In particular, let us assume that Person 3 is a friend of Persons 1 and 2. That is, 
\[
\mathcal{N}_{1}=\left\{ 2,3\right\},\quad \mathcal{N}_{2}=\left\{1,3\right\},\quad \mathcal{N}_{3}=\left\{ 1,2\right\},\quad \mathcal{N}_{4}=\left\{ 5\right\},\quad \mathcal{N}_{5}=\left\{ 4\right\}.
\]
In this case, the network becomes undirected. Repeating the previous exercise, the new network generates the marginal equilibrium distributions depicted in Table~\ref{table: marg prob 2}. 
\begin{table}
\centering
\begin{threeparttable}
\centering
\caption{Marginal Equilibrium Probabilities for Undirected Network}\label{table: marg prob 2}
\begin{tabular}{|c|c|c|c|c|}
\hline
$\mu_{1}\left( o\right) =0.12$ & $\mu_{2}\left( o\right) =0.12$ & $\mu
_{3}\left( o\right) =0.12$ & $\mu_{4}\left( o\right) =0.30$ & $\mu
_{5}\left( o\right) =0.30$ \\ 
$\mu_{1}\left( 1\right) =0.21$ & $\mu_{2}\left( 1\right) =0.42$ & $\mu
_{3}\left( 1\right) =0.22$ & $\mu_{4}\left( 1\right) =0.50$ & $\mu
_{5}\left( 1\right) =0.50$ \\ 
$\mu_{1}\left( 2\right) =0.67$ & $\mu_{2}\left( 2\right) =0.46$ & $\mu
_{3}\left( 2\right) =0.66$ & $\mu_{4}\left( 2\right) =0.20$ & $\mu
_{5}\left( 2\right) =0.20$ \\ \hline
\end{tabular}%
\vspace{1ex}
\begin{tablenotes}
\item {\footnotesize Notes: These probabilities are estimated using simulated data, sample size $=15000$.}
\end{tablenotes}
\end{threeparttable}
\end{table}
From these marginals, the probabilities of mistakes are presented on Table~\ref{table: prob of mistakes 2}. 
\begin{table}
\centering
\begin{threeparttable}
\centering
\caption{Probabilities of Mistakes for Undirected Network}\label{table: prob of mistakes 2}
\begin{tabular}{|c|c|c|c|c|c|}
\hline
& Person 1 & Person 2 & Person 3 & Person 4 & Person 5 \\ \hline
Probability of Mistakes & 33\% & 58\% & 34\% & 50\% & 50\% \\ \hline
\end{tabular}%
\vspace{1ex}
\begin{tablenotes}
\item {\footnotesize Notes: These probabilities are computed using the estimated marginal equilibrium probabilities from Table~\ref{table: marg prob 2}.}
\end{tablenotes}
\end{threeparttable}
\end{table}
The probabilities of making mistakes decrease for Persons 1, 2, and 3. But notice that the change is larger for Persons 1 and 3 as they share the same preferences over the alternatives.{\normalsize\hfill }$\blacksquare $\

\section{Empirical Content of the Model\label{MI}}
\noindent This section provides conditions under which the researcher can uniquely recover (from a long sequence of choices) the set of connections $\Gamma =\left( \mathcal{A},e\right)$, the profile of strict preferences $\succ=\left( \succ_{a}\right)_{a\in \mathcal{A}}$, the attention mechanism $\operatorname{Q}=\left( \operatorname{Q}_{a}\right)_{a\in \mathcal{A}}$, and the rates of the Poisson alarm clocks $\lambda=\left( \lambda_{a}\right)_{a\in \mathcal{A}}$.

We separate the identification analysis in two parts. Let $\operatorname{P}=\left( \operatorname{P}_{a}\right)_{a\in \mathcal{A}}$ be the profile of choice probabilities of people in the network. Each $\operatorname{P}_{a}\left( v \mid \mathbf{y}\right) :\overline{\mathcal{Y}}\times\overline{\mathcal{Y}}^{A}\rightarrow\left( 0,1\right)$ specifies the (ex-ante) probability that Person $a$ selects option $v$ when the choice configuration is $\mathbf{y}$. Recall that, in our model,
\begin{eqnarray*}
\operatorname{P}_{a}\left( v \mid \mathbf{y}\right) =\operatorname{Q}_{a}\left( v \mid \mathbf{y}\right) \prod\nolimits_{v'\in \mathcal{Y}, v' \succ_{a} v}\left( 1-\operatorname{Q}_{a}\left( v' \mid \mathbf{y}\right) \right),
\end{eqnarray*}
and the probability of selecting the default option $o$ is $\prod\nolimits_{v\in\mathcal{Y}}\left( 1-\operatorname{Q}_{a}\left( v \mid \mathbf{y}\right)
\right)$. 
First, we show that each set of conditional choice probabilities $\operatorname{P}$ maps into a different set of connections, profile of strict preferences, and attention mechanism. Thus, knowledge of the set of the conditional choice probabilities allows us to uniquely recover all the elements of the model. Second, we build identification of the conditional choice probabilities $\operatorname{P}$ from a long sequence of choices.

\subsection{Identification of the Model from \texorpdfstring{$\operatorname{P}$}{the Conditional Choice Probabilities}}\label{PIP}

\noindent Under Assumptions A1 and A2, changes in the choices of friends induce stochastic variation of the consideration sets. Assumption A3 guarantees this variation is monotonic in the sense that the probability of considering one option increases with the number of friends that are currently adopting it. This stochastic variation in choices allows us to recover the set of connections between the people in the network and the ranking of preferences of each of them. We then sequentially identify the attention mechanism of each person moving from the most preferred alternative to the least preferred one. Proposition~\ref{ID1} presents our first identification result.

\begin{proposition}\label{ID1}
Under Assumptions A1-A3, the set of connections $\Gamma$, the profile of strict preferences $\succ$, and the attention mechanism $\operatorname{Q}$ are point identified from $\operatorname{P}$.
\end{proposition}

The next example sheds some light on the identification strategy in Proposition~\ref{ID1}.\bigskip

\noindent \textbf{Example 3.} Suppose there are three people $\mathcal{A}=\left\{ 1,2,3\right\}$ that select between two alternatives $\mathcal{Y}=\left\{ 1,2\right\} $ and the default option $o$. The researcher knows $\operatorname{P}_{1}$, $\operatorname{P}_{2}$, and $\operatorname{P}_{3}$. Let us consider Person 1. Let $\mathbf{y}$ be such that $y_1=o$. The probability that Person 1 selects the default option $o$ (given a profile of choices $\mathbf{y}$ with $y_1=o$) is 
\begin{equation*}
\operatorname{P}_{1}\left( o \mid \mathbf{y}\right) =\left( 1-\operatorname{Q}_{1}\left( 1 \mid o,\operatorname{N}_{1}^{1}\left( \mathbf{y}\right)\right) \right) \left( 1-\operatorname{Q}_{1}\left( 2 \mid o,\operatorname{N}_{1}^{2}\left( \mathbf{y}\right) \right) \right).
\end{equation*}
Under A3, we get that $2\in \mathcal{N}_{1}$ if and only if 
\begin{equation*}
\operatorname{P}_{1}\left( o \mid o,o,o\right) >\operatorname{P}_{1}\left( o \mid o,1,o\right) .
\end{equation*}
Indeed, if $2\in \mathcal{N}_{1}$, then the probability of choosing the default option by Person 1 should decrease if Person $2$ picks something else. Also, if $2\not\in \mathcal{N}_{1}$, then the probability of choosing the default option by Person 1 should be invariant to choices of Person 2. Similarly, $3\in \mathcal{N}_{1}$ if and only if $\operatorname{P}_{1}\left( o \mid o,o,o\right) >\operatorname{P}_{1}\left( o \mid o,o,1\right)$. Thus, we can learn the set of friends of Person 1 from observed $\operatorname{P}_{1}$. Let us assume that $\mathcal{N}_{1}=\left\{ 2\right\}$. To recover the preferences of Person 1 note that 
\begin{equation*}
\begin{array}{lcc}
\operatorname{P}_{1}\left( 1 \mid \mathbf{y}\right) =\operatorname{Q}_{1}\left( 1 \mid o,\operatorname{N}_{1}^{1}\left( \mathbf{y}\right) \right) & \text{if} & 1\succ_{1}2 \\ 
\operatorname{P}_{1}\left( 1 \mid \mathbf{y}\right) =\operatorname{Q}_{1}\left( 1 \mid o,\operatorname{N}_{1}^{1}\left( \mathbf{y}\right) \right) \left( 1-\operatorname{Q}_{1}\left( 2 \mid o,\operatorname{N}_{1}^{2}\left( \mathbf{y}\right) \right) \right) & \text{if} & 2\succ_{1}1
\end{array}
.
\end{equation*}
Thus, $2\succ_{1}1$ if and only if 
\begin{equation*}
\operatorname{P}_{1}\left( 1 \mid o,o,o\right) >\operatorname{P}_{1}\left( 1 \mid o,2,o\right) .
\end{equation*}
Suppose that, indeed, we get that $2\succ_{1}1$. We can finally recover the attention mechanism (for $y_{_{1}}=o$) via the next four probabilities in the data 
\begin{equation*}
\begin{array}{ll}
\operatorname{P}_{1}\left( 2 \mid o,o,o\right) =\operatorname{Q}_{1}\left( 2 \mid o,0\right) & \operatorname{P}_{1}\left( 2 \mid o,2,o\right) =\operatorname{Q}_{1}\left( 2 \mid o,1\right) \\ 
\operatorname{P}_{1}\left( 1 \mid o,o,o\right) =\operatorname{Q}_{1}\left( 1 \mid o,0\right) \left( 1-\operatorname{Q}_{1}\left( 2 \mid o,0\right)\right) &\operatorname{P}_{1}\left( 1 \mid o,1,o\right) =\operatorname{Q}_{1}\left( 1 \mid o,1\right) \left( 1-\operatorname{Q}_{1}\left( 2 \mid o,0\right)\right)
\end{array}
.
\end{equation*}
By considering two other choice profiles $\mathbf{y}$ with $y_1=1$ and $y_1=2$ (instead of $y_1=o$), respectively, we can fully recover the attention mechanism of Person 1. By a similar exercise we can recover the sets of friends, preferences, and the attention mechanisms for Persons 2 and 3. {\normalsize \hfill }$\blacksquare $

\subsection{Identification of \texorpdfstring{$\operatorname{P}$}{the Conditional Choice Probabilities}}
\noindent This section studies identification of the conditional choice probabilities, $\operatorname{P}$, and the rates of the Poisson alarm clocks from two different datasets. These two datasets coincide in that they contain long sequences of choices from people in the network. They differ in the timing at which
the researcher observes these choices. In Dataset 1 people's choices are observed in real time. This allows the researcher to record the precise moment at which a person revises her strategy and the configuration of choices at that time. In Dataset 2 the researcher simply observes the joint configuration of choices at fixed time intervals.

Let us assume the researcher observes people's choices at time intervals of length $\Delta$ and can consistently estimate $\Pr\left(\mathbf{y}^{t+\Delta }=\mathbf{y}'\mid\mathbf{y}^{t}=\mathbf{y}\right)$ for each pair $\mathbf{y}',\mathbf{y}\in\overline{\mathcal{Y}}^{A}$. We will capture these transition probabilities by a matrix $\mathcal{P}\left( \Delta \right)$. (Here again, we will assume that the choice configurations are ordered according to the lexicographic order when we construct $\mathcal{P}\left(\Delta \right)$.) The connection between $\mathcal{P}\left( \Delta \right)$ and the transition rate matrix $\mathcal{M}$ described in Equation (\ref{T}) is given by 
\begin{equation*}
\mathcal{P}\left( \Delta \right) =e^{\left( \Delta \mathcal{M}\right) },
\end{equation*} 
where $e^{\left( \Delta \mathcal{M}\right)}$ is the matrix exponential of $\Delta \mathcal{M}$.
The two datasets we consider differ regarding $\Delta$: in Dataset 1 we let the time interval be very small. This is an ideal dataset that registers people's choices at the exact time in which any given person revises her choice. As we mentioned earlier, with the proliferation of online platforms and scanner this sort of data might indeed be available for some applications. In Dataset 2 we allow the time interval to be of arbitrary size. The next table formally describes Datasets 1 and 2 
\begin{equation*}
\begin{array}{ll}
\text{\textbf{Dataset 1}} & \text{The researcher knows }\lim_{\Delta
\rightarrow 0}\mathcal{P}\left( \Delta \right) \\ 
\text{\textbf{Dataset 2}} & \text{The researcher knows }\mathcal{P}\left(
\Delta \right)
\end{array}
\end{equation*}
In both cases, the identification question is whether (or under what extra restrictions) it is possible to uniquely recover $\mathcal{M}$ from $\mathcal{P}\left( \Delta \right)$. The first result in this section is as follows.

\begin{proposition}[Dataset 1]\label{ID2}
The conditional choice probabilities $\operatorname{P}$ and the rates of the Poisson alarm clocks $\lambda$ are identified from Dataset 1.
\end{proposition}

The proof of Proposition~\ref{ID2} relies on the fact that when the time interval between the observations goes to zero, then we can recover $\mathcal{M}$. There are at least two well-known cases that produce the same outcome without assuming $\Delta \rightarrow 0$. One of them requires the length of the interval $\Delta$ to be below a threshold $\overline{\Delta }$. The main difficulty of this identification approach is that the value of the threshold depends on the details of the model that are unknown to the researcher. The second case requires the researcher to observe the dynamic system at two different intervals $\Delta_{1}$ and $\Delta_{2}$ that are not multiples of each other. (See, e.g., \citealp{blevins2017identifying}, and the literature therein.)

The next proposition states that, by adding an extra restriction, the transition rate matrix can be identified from people's choices even if these choices are observed at the endpoints of discrete time intervals. In this case, the researcher needs to know the rates of the Poisson alarm clocks or normalize them in empirical work.

\begin{proposition}[Dataset 2]\label{ID3}
If Assumption A2 is satisfied, the researcher knows $\lambda$, and $\mathcal{M}$ has distinct eigenvalues that do not differ by an integer multiple of $2\pi i/\Delta $, where $i$ here denotes the imaginary unit, then the conditional choice probabilities $\operatorname{P}$ are generically identified from Dataset 2.
\end{proposition}

The key element in proving Proposition~\ref{ID3} is that the transition rate matrix in our model is rather parsimonious. To see why, recall that, at any given time, only one person revises her selection with nonzero probability. This feature of the model translates into a transition rate matrix $\mathcal{M}$ that has many zeros in known locations (see Example 1). It is important to remark that this aspect of the model is not shared by models of discrete-time revision processes, where people in the network revise choices simultaneously at fixed time intervals.

\section{Extensions of the Model}\label{E}

\subsection{Random Preferences}\label{RP}

\noindent This section extends the initial model to allow randomness in preferences and in consideration sets. In this case, the choice rule $\operatorname{R}_{a}\left( \cdot  \mid \mathcal{C}\right)$ from Section~\ref{M} is not an indicator function but a distribution on $\mathcal{Y}$. We naturally let $\operatorname{R}_{a}\left( v \mid \mathcal{C}\right) =0$ if $v\notin \mathcal{C}$. Note that the relation between preferences and consideration sets is completely unrestricted.

Keeping unchanged the other parts of the model, the probability that Person $a$ selects (at the moment of choosing) alternative $v\in \mathcal{Y}$ is given by 
\begin{equation}
\operatorname{P}_{a}\left( v \mid \mathbf{y}\right) =\sum\nolimits_{ 
\mathcal{C\subseteq }2^{\mathcal{Y}}}\operatorname{R}_{a}\left( v \mid \mathcal{C}
\right) \prod\nolimits_{v'\in \mathcal{C}}\operatorname{Q}_{a}\left(
v' \mid y_{a},\operatorname{N}_{a}^{v'}\left( \mathbf{y}
\right) \right) \prod\nolimits_{v'\notin \mathcal{C}}\left( 1- 
\operatorname{Q}_{a}\left( v' \mid y_{a},\operatorname{N}_{a}^{v'}\left( \mathbf{y}\right) \right) \right).  \label{Gy}
\end{equation}
The probability of selecting the default option $o$ is (as before) $\prod\nolimits_{v\in \mathcal{Y}}\left( 1-\operatorname{Q}_{a}\left( v \mid y_{a}, \operatorname{N}_{a}^{v}\left( \mathbf{y}\right) \right) \right)$. The next example illustrates the random choice rule with the well-known logit model.\bigskip

\noindent \textbf{Example 4.} If we use the logit model to represent the random preferences of Person $a$, then the probability that the person
selects alternative 1 when alternative 2 is also part of her consideration set would be given by 
\begin{equation*}
\operatorname{R}_{a}\left( 1 \mid \left\{ 1,2\right\} \right) =\frac{\exp\left( \mathrm{U}_{a}^{1}\right) }{\exp \left( \mathrm{U}_{a}^{1}\right) +\exp
\left( \mathrm{U}_{a}^{2}\right) }.
\end{equation*}%
In this expression, $\mathrm{U}_{a}^{1}$ and $\mathrm{U}_{a}^{2}$ are the mean expected utilities that Person $a$ gets from alternatives 1 and 2, respectively. 
{\normalsize \hfill }$\blacksquare $\bigskip

Under this variant of the initial model, the identification of $\operatorname{P}$ follows directly from Propositions~\ref{ID2} and~\ref{ID3}. We will thereby focus on recovering the set of connections, the choice rule, and the attention mechanism from $\operatorname{P}$. The main result is as follows.

\begin{proposition}
\label{IDE1}Suppose Assumptions A1-A3 are satisfied. Then, the set of connections $\Gamma$ and the attention mechanism $\operatorname{Q}$ are point identified from $\operatorname{P}$. For each $a\in \mathcal{A}$, the random preferences $\operatorname{R}_{a}$ are also point identified if, and only if, in addition, we have that $\left\vert \mathcal{N}_{a}\right\vert\geq Y-1$.
\end{proposition}

\noindent \textbf{Remark. }The last result extends to the case in which the random preferences include the default option $o$ with only one caveat. In
this case, the attention mechanism can be recovered up to ratios of the form $\operatorname{Q}_{a}\left( v \mid y_{a},\operatorname{N}_{a}^{v}\left( \mathbf{y}\right)\right) /\operatorname{Q}_{a}\left( v \mid y_{a},0\right)$. That is, we can only recover how much \textit{extra} attention a person pays to each option as more of her friends select that option.\bigskip

As in our previous results, under Assumptions A1-A3, observed variation in the choices of friends induce stochastic variation of the consideration sets and this variation suffices to recover the connections between the people in the network and the attention mechanism. The only difference with respect to the case of deterministic preferences is that with random preferences we need a larger number of friends for each person, i.e., $\left\vert \mathcal{N}_{a}\right\vert \geq Y-1$ for each $a\in \mathcal{A}$. The extra condition guarantees the matrix of coefficients for the $\operatorname{R}_{a}'$s in expression (\ref{Gy}) is full column rank. Indeed, we show that $\left\vert\mathcal{N}_{a}\right\vert \geq Y-1$ is not only sufficient, but necessary, to this end. We illustrate the last result by a simple extension of Example 3 above.\bigskip

\noindent \textbf{Example 3 (continued).} Let us keep all the structure of Example 3 except for people's preferences, which we now assume are random. The identification of the set of connections and the attention mechanism follows from similar ideas. Thus, we will only focus on recovering $\operatorname{R}_{1}$, $\operatorname{R}_{2}$, and $\operatorname{R}_{3}$. Consider the following system of equations for Person 1  that connects the observed $\operatorname{P}$ with $\operatorname{Q}$ and $\operatorname{R}$
\begin{equation*}
\left( 
\begin{array}{c}
\operatorname{P}_{1}\left( 1 \mid y_{1},o,o\right) /\operatorname{Q}_{1}\left( 1 \mid y_{1},0\right) \\ 
\operatorname{P}_{1}\left( 1 \mid y_{1},2,o\right) /\operatorname{Q}_{1}\left( 1 \mid y_{1},0\right)
\end{array}
\right) =\left( 
\begin{array}{cc}
1-\operatorname{Q}_{1}\left( 2 \mid y_{1},0\right) & \operatorname{Q}_{1}\left( 2 \mid y_{1},0\right) \\ 
1-\operatorname{Q}_{1}\left( 2 \mid y_{1},1\right) & \operatorname{Q}_{1}\left( 2 \mid y_{1},1\right)
\end{array}
\right) \left( 
\begin{array}{c}
\operatorname{R}_{1}\left( 1 \mid \left\{ o,1\right\} \right) \\ 
\operatorname{R}_{1}\left( 1 \mid \left\{ o,1,2\right\} \right)
\end{array}
\right).
\end{equation*}
The fact that $\operatorname{R}_{1}\left( 1 \mid \left\{ 1\right\} \right)$ and $\operatorname{R}_{1}\left( 1 \mid \left\{ 1,2\right\} \right)$ can be recovered follows because, by Assumption A3, we have that the determinant of the matrix of the coefficients in the above system of equations is different from zero:
\begin{equation*}
\det \left( 
\begin{array}{cc}
1-\operatorname{Q}_{1}\left( 2 \mid y_{1},0\right) & \operatorname{Q}_{1}\left( 2 \mid y_{1},0\right) \\ 
1-\operatorname{Q}_{1}\left( 2 \mid y_{1},1\right) & \operatorname{Q}_{1}\left( 2 \mid y_{1},1\right)
\end{array}
\right) =\operatorname{Q}_{1}\left( 2 \mid y_{1},1\right) -\operatorname{Q}_{1}\left( 2 \mid y_{1},0\right) >0.
\end{equation*}
The extra condition, $\left\vert \mathcal{N}_{a}\right\vert \geq Y-1$, and Assumption A3 guarantee that the matrix of coefficients for the $\operatorname{R}_{a}$'s is always full column rank. {\normalsize \hfill }$\blacksquare $

\subsection{No Default Option\label{default}}

\noindent In the initial model, the default option plays a special role: it is chosen if, and only if, nothing else is considered. In some settings, such default option may not exist.\footnote{Also, see \citet{horan2019random}.} This section offers a variant of the model that accommodates to this possibility.\footnote{For alternative ways to close the model see, for example, \citet{barseghyan2019discrete}.}

Let us assume there is no default option $o$, so that $\overline{\mathcal{Y}}=\mathcal{Y}$. The formation process of the consideration set is as before except that, since there is no default option, we need to specify what people do when the consideration set is empty. Given the dynamic nature of our model, we will simply assume that each person sticks to her previous choice if no alternative receives further consideration. Formally, the probability that Person $a$ selects (at the moment of choosing) alternative $v\in\mathcal{Y}$ is given by 
\begin{equation}
\operatorname{P}_{a}\left( v \mid \mathbf{y}\right) =\operatorname{Q}_{a}\left( v%
 \mid \mathbf{y}\right) \prod\nolimits_{v'\in \mathcal{Y}%
,v'\succ_{a}v}\left( 1-\operatorname{Q}_{a}\left( v' \mid \mathbf{y}\right) \right) +1(v=y_{a})\prod\nolimits_{v'\in \mathcal{Y}}\left( 1-\operatorname{Q}_{a}\left( v' \mid \mathbf{y} \right)\right).
\end{equation}

\begin{proposition}\label{IDE2}
Suppose that Assumptions A1-A3 are satisfied and $Y \geq 3$. Then, the set of connections $\Gamma$, the profile of strict preferences $\succ$, and the attention mechanism $\operatorname{Q}$ are point identified from $\operatorname{P}$.
\end{proposition}

\subsection{More General Peer Effects in Consideration Sets}\label{subsec: general consideration}

\noindent In our initial model, the probability that Person $a$ faces consideration set $\mathcal{C}$ given a choice configuration $\mathbf{y}$ takes the form of
\begin{equation*}
\prod\nolimits_{v\in \mathcal{C}}\operatorname{Q}_{a}\left( v \mid y_{a},
\operatorname{N}_{a}^{v}\left( \mathbf{y}\right) \right) \prod\nolimits_{v\notin 
\mathcal{C}}\left( 1-\operatorname{Q}_{a}\left( v \mid y_{a},\operatorname{N}_{a}^{v}\left( \mathbf{y}\right) \right) \right) .
\end{equation*}%
This approach entails a multiplicative separable specification of alternatives in the consideration sets. It also assumes that the probability of considering an option depends only on the total number of friends who pick that option, but not on the identity of these friends (i.e., the effect of friends' choices is symmetric across friends). We next show that neither of these assumptions is essential for our approach. Indeed, all previous insights can be used here to extend the initial results. 

For each Person $a$ and configuration $\mathbf{y}$, let $\eta_{a}(\cdot \mid \mathbf{y})$ be an attention index function from $2^{\mathcal{Y}}$ to the positive reals. The value $\eta_a(\mathcal{C} \mid \mathbf{y})$ captures the attention that Person $a$ pays to the set of alternatives $\mathcal{C}\in2^{\mathcal{Y}}$ given the choice configuration $\mathbf{y}$. The attention-index measures how enticing a consideration set is (see \citet{aguiar2019doesrandom} for further details). We define the probability of facing consideration set $\mathcal{C}$ as 
\begin{equation*}
\dfrac{\eta_a(\mathcal{C} \mid \mathbf{y})}{\displaystyle\sum\nolimits_{\mathcal{D}\subseteq\mathcal{Y}}\eta_a(\mathcal{D} \mid \mathbf{y})}.
\end{equation*}
In this model, the consideration set probabilities are as in \citet{brady2016menu}. Since $\eta_a$ can be identified only up to scale, we normalize the attention-index for the empty set to be 1 (i.e., $\eta(\emptyset \mid \mathbf{y})=1$ for all $\mathbf{y}\in \overline{\mathcal{Y}}^{A}$).\footnote{If one instead normalizes $\sum\nolimits_{\mathcal{D}\subseteq\mathcal{Y}}\eta_a(\mathcal{D} \mid \mathbf{y})=1$, then we get the consideration rule of \citet{aguiar2017random}.} This more general setting covers our initial model, which is based on \citet{manzini2014stochastic}, as a special case.
\par
By combining preferences and this specification of stochastic consideration sets, the probability that Person $a$ selects (at the moment of choosing) alternative $v\in \mathcal{Y}$ is given by 
\begin{equation*}
\operatorname{P}_{a}\left( v \mid \mathbf{y}\right) =\sum\nolimits_{\mathcal{C}\in 2^{\mathcal{Y}}:v\in \mathcal{C}} \operatorname{R}_{a}\left( v \mid \mathcal{C}\right)\dfrac{\eta_{a}\left( \mathcal{C} \mid  \mathbf{y}\right)}{\displaystyle\sum\nolimits_{\mathcal{D}\subseteq\mathcal{Y}}\eta_a(\mathcal{D} \mid \mathbf{y})}.
\end{equation*}
The probability of selecting the default option $o$ is just $\eta_{a}\left( \emptyset \mid \mathbf{y}\right)/\sum\nolimits_{\mathcal{D}\subseteq\mathcal{Y}}\eta_a(\mathcal{D} \mid \mathbf{y})$. To better understand the connection between our initial model and this extension note that, for any $v\in\mathcal{Y}$ and $\mathcal{C}\subseteq\mathcal{Y}$ such that $v\not\in\mathcal{C}$, we have that
\[
\dfrac{\eta_a(\mathcal{C}\cup\{v\} \mid \mathbf{y})}{\eta_a(\mathcal{C}\mid \mathbf{y})}=\dfrac{\operatorname{Q}_a(v\mid \mathbf{y})}{1-\operatorname{Q}_a(v\mid \mathbf{y})}.
\]
Thus,
\[
\operatorname{Q}_a(v\mid \mathbf{y})=\dfrac{\eta_a(\mathcal{C}\cup\{v\} \mid \mathbf{y})}{\eta_a(\mathcal{C}\cup\{v\} \mid \mathbf{y})+\eta_a(\mathcal{C}\mid \mathbf{y})}.
\]
Based on this alternative specification, we accommodate Assumptions A1 and A3 as follows.

\bigskip
\noindent \textbf{(A1')} For each $a\in \mathcal{A}$, $v\in\mathcal{Y}$, and $\mathbf{y}\in \overline{\mathcal{Y}}^{A}$, there exists $\mathcal{C}\in 2^\mathcal{Y}$ such that $v\succ_a v'$ for all $v'\in\mathcal{C}$ and $\eta_{a}\left(\mathcal{C}\cup\{v\} \mid \mathbf{y}\right)>0$.
\medskip

\noindent \textbf{(A3')} For each $a\in \mathcal{A}$, $\mathcal{C}\in 2^\mathcal{Y}$, and 
$\mathbf{y},\mathbf{y}^*\in \overline{\mathcal{Y}}^{A}$, such that $\mathbf{y}$ is different from $\mathbf{y}^*$ just in one component $a^*$,
\begin{enumerate}
    \item $a^*\not\in \mathcal{N}_a \text{ or } y_{a^*},\:y^*_{a^*}\not\in\mathcal{C} \implies\eta_{a}\left( \mathcal{C} \mid  \mathbf{y}\right)=\eta_{a}\left( \mathcal{C} \mid  \mathbf{y}^*\right)$;
    \item $a^*\in \mathcal{N}_a,\: y_{a^*}\in\mathcal{C} \text{ and } y^*_{a^*}\not\in\mathcal{C}\implies\eta_{a}\left(\mathcal{C} \mid  \mathbf{y}\right)>\eta_{a}\left( \mathcal{C} \mid  \mathbf{y}^*\right)$.
\end{enumerate}

Assumption A3'(i) states that the attention a person pays to a given set is invariant to the choices of those who are not connected with the person and to alternatives that do not enter the set. Assumption A3'(ii) means that switches of friends to a new alternative boost the attention for all sets that contain this new alternative. Note that Assumption A3' does not assume that different friends affect consideration probabilities symmetrically. That is, the model allows the possibility that some friends have a bigger effect than others. Since the probability of facing consideration set $\mathcal{C}$ is proportional to the inverse of the total attention, $\sum\nolimits_{\mathcal{D}\subseteq\mathcal{Y}}\eta_a(\mathcal{D} \mid \mathbf{y})$, even if peers are switching between alternatives that are not elements of $\mathcal{C}$, the probability of facing $\mathcal{C}$ may change.

The next proposition states that, with this general alternative specification, the network structure and the profile of preferences can be uniquely recovered from the conditional choice probabilities. Though the attention mechanism is just partially-identified, it is point identified under additional restrictions that reduce the dimensionality of the problem; we describe some of these restrictions in the next result. 

\begin{proposition}\label{IDE3}
Suppose that Assumptions A1', A2, and A3' are satisfied. Then, the set of connections $\Gamma$ and the profile of strict preferences $\succ$ are point identified from $\operatorname{P}$. If additionally for any $a\in\mathcal{A}$, $\mathbf{y}\in \overline{\mathcal{Y}}^{A}$, and $\mathcal{C},\mathcal{D}\in 2^{\mathcal{Y}}$ one of the following holds
\begin{enumerate}
    \item \citep{manzini2014stochastic} $\eta_{a}(\mathcal{C} \mid \mathbf{y})=\displaystyle\prod\nolimits_{v\in\mathcal{C}}\eta_{a}(\{v\} \mid \mathbf{y})$;
    \item \citep{dardanoni2020inferring}
    $\left\vert\mathcal{C}\right\vert=\left\vert\mathcal{D}\right\lvert\implies \eta_{a}(\mathcal{C} \mid \mathbf{y})=\eta_{a}(\mathcal{D} \mid \mathbf{y})$;
    \item $\eta_{a}(\mathcal{C} \mid \mathbf{y})=\displaystyle\sum\nolimits_{v\in\mathcal{C}}\eta_{a}(\{v\} \mid \mathbf{y})$;
    \item $\eta_{a}(\mathcal{C} \mid \mathbf{y})=\eta_{a}(\{v^*\} \mid \mathbf{y})$, where $v^*\in\mathcal{C}$ satisfies $v^*\succ_a v'$ for all $v'\in \mathcal{C}$;
\end{enumerate}
then $\{\eta_a\}_{a\in\mathcal{A}}$ is also pointidentified.
\end{proposition}
Condition (i) in Proposition~\ref{IDE3} restates our main result using the consideration set formation model of \citet{manzini2014stochastic} ---see also \citet{manski1977structure}. Condition (ii) shows that the model analyzed in \citet{dardanoni2020inferring} ---where the sets of equal cardinality are considered with equal probability--- also imposes enough restrictions to recover consideration probabilities. Conditions (iii) and (iv) are new. Condition (iii) is similar to condition (i), but defines ``aggregate'' attention as a sum of singleton-attentions. Condition (iv) postulates that consideration probability of a set only depends on the best alternative in that set. 

\section{Estimation of the Model Parameters}\label{RR}
\noindent This section provides an estimator of the model parameters that uses discrete-time data (Dataset 2) and assesses its quality in Monte Carlo simulations. 
\par
Let $\theta =\left( \Gamma ,\succ,\operatorname{Q}\right)$ be an element of the space of possible parameters we want to estimate. (We normalize the intensity parameter $\lambda_{a}$ to $1$ for all $a\in\mathcal{A}$.) Note that since the number of people in the network and the number of choices is finite, the parameter space for $\Gamma$ and $\succ$ is a finite set.  For each $\theta$ we can construct the transition rate matrix $\mathcal{M}\left( \theta \right)$ using Equations (\ref{a}) and (\ref{T}). In turn, this information allows us to calculate the transition matrix 
\begin{equation*}
\mathcal{P}\left( \theta ,\Delta \right) =e^{\Delta \mathcal{M}\left( \theta\right) }.
\end{equation*}
Given a sample of network configurations $\{\mathbf{y}_{t}\}_{t=0}^T$, we can use the latter to build the log-likelihood function $\mathrm{L}_{T}\left(\theta \right) =\Sigma_{t=0}^{T-1}\ln \mathcal{P}_{\iota \left( \mathbf{y}_{t}\right),\iota \left(\mathbf{y}_{t+1}\right) }\left( \theta ,\Delta\right)$, where $\iota \left(\mathbf{y}\right) \in \left\{ 1,2,\dots,\overline{\mathcal{Y}}^{A}\right\}$ is the position of $\mathbf{y}$ according the lexicographic order, and $\mathcal{P}_{k,m}\left( \theta,\Delta \right)$ is the $(k,m)$-th element of the matrix $\mathcal{P}\left(\theta ,\Delta \right)$. Finally, the maximum likelihood estimator of the true parameter value can be defined as 
\begin{equation*}
\widehat{\theta}_{T}=\arg \max\nolimits_{\theta }\operatorname{L}_{T}\left( \theta
\right).
\end{equation*}
\par
To evaluate the finite-sample properties of our estimator we conducted several Monte Carlo experiments. In all experiments we simulated discrete-time data from the specification presented in Example 2 in Section~\ref{Equi}. In particular, we assumed that
\[
2\succ_{1}1,\quad 1\succ_{2}2,\quad 2\succ_{3}1,\quad 1\succ_{4}2,\quad 1\succ_{5}2,
\]
\[
\operatorname{Q}\left( v \mid 0\right) =\frac{1}{4},\quad\operatorname{Q}\left( v \mid 1\right) =\frac{3}{4},\quad\operatorname{Q}\left( v \mid 2\right) =\frac{7}{8},
\]
and
\[
\mathcal{N}_{1}=\left\{ 2,3\right\},\quad \mathcal{N}_{2}=\left\{1,3\right\},\quad \mathcal{N}_{3}=\left\{ 1,2\right\},\quad \mathcal{N}_{4}=\left\{ 5\right\},\quad \mathcal{N}_{5}=\left\{ 4\right\}.
\]
\par
First, we estimate $\operatorname{Q}$ under the assumption that the network structure and preferences are known. The experiment was replicated 1000 times for 6 different sample sizes. The results of these simulations are presented in Table~\ref{table: bias and RMSE}. The estimator of $\operatorname{Q}$ performs well in terms of the mean bias and the root mean squared error. As expected, the bias and the root mean squared error decrease with the sample size. (See Appendix~\ref{app: simul} for more details on how the data was generated.)
\begin{table}
\centering
\begin{threeparttable}
\centering
\caption{Bias and Root Mean Squared Error (RMSE) ($\times 10^{-3}$)}\label{table: bias and RMSE}
\begin{tabular}{clcccccc}
\hline
\hline
&&\multicolumn{6}{c}{Sample Size}\\
\cmidrule{3-8}
Attention Probabilities                     &       & 10  & 50    & 100   & 500  & 1000  & 5000 \\ 
\hline
$\operatorname{Q}\left( v \mid 0\right)$    & Bias  &213.6 & 164.9 & 133.2 & 65.1 & 42.3 & 9.4. \\
                                            & RMSE  &259.1 & 172.9 & 137.4 & 66.6 & 43.3 & 10.2 \\ 
$\operatorname{Q}\left( v \mid 1\right)$    & Bias  &66.4  & 53    & 46.9  & 27.2 & 18   & 2.7  \\ 
                                            & RMSE  &103.9 & 66.9  & 55.6  & 31   & 20.9 & 5.6  \\ 
$\operatorname{Q}\left( v \mid 2\right)$    & Bias  &53.8  & 49.8  & 42.9  & 23.6 & 15.1 & 0.3  \\ 
                                            & RMSE  &94.3  & 64.6  & 52.8  & 27.2 & 17.8 & 4.4  \\ 
\hline
\end{tabular}
\vspace{1ex}
\begin{tablenotes}
\item {\footnotesize Notes: The sample sizes of $10$, $50$, $100$, $500$, $1000$, and $5000$ correspond to $\Delta$ equal to $2500$, $500$, $250$, $50$, $25$, and $5$. The number of replications is $1000$.}
\end{tablenotes}
\end{threeparttable}
\end{table}

Next we estimate the whole parameter vector since our method allows to consistently estimate the network structure and the preference orders of individuals as well. Since in our example, without any restrictions, there are $2^{A(A-1)}=1,048,576$ possible networks and $(Y!)^{A}=32$ strict preference orders, to make the problem computationally tractable we restricted the parameter space for $\Gamma$ by making the following assumptions: (i) each person has at most two friends; (ii) the attention mechanism is invariant across people and alternatives; and (iii) the network is undirected. As a result, the number of possible networks becomes $112$ (the number of possible preference orders is still $32$). The experiment was conducted $500$ times for different sample sizes. Table~\ref{table: network and pref} presents the results of these simulations. With just $50$ observations the network structure is correctly estimated $94.4$ percent times. For the sample size of $500$ the network structure and the preferences are correctly estimated in all simulations.
\begin{table}[h]
\centering
\begin{threeparttable}
\centering
\caption{Correctly Estimated Network \& Preferences}\label{table: network and pref}
\begin{tabular}{ccccc}
\hline
\hline
Sample Size             & 10    & 50    & 100   & 500   \\ 
\hline
Network                 & 32.4\%  & 94.4\%  & 99.8\% & 100\% \\
Preferences             & 34.6\%  & 85.6\%  & 97.6\%  & 100\% \\
Network \& Preferences  & 13.4\%  & 83\%  & 97.4\%  & 100\% \\
\hline
\end{tabular}
\vspace{1ex}
\begin{tablenotes}
\item {\footnotesize Notes: The sample sizes of $10$, $50$, $100$, and $500$ correspond to $\Delta$ equal to $2500$, $500$, $250$, and $50$. The number of replications is $500$.}
\end{tablenotes}
\end{threeparttable}
\end{table}

\section{Empirical Application}\label{sec: empitrical application}

\noindent In this section, we apply our framework to an experimental dataset used by \citet{bai2019predicting} to test models that aim to predict the visual focus of attention of people. In each experiment, a group of people were asked to play the Resistance game. This is a party game in which people communicate with each other in order to find the deceptive players.\footnote{See \url{https://en.wikipedia.org/wiki/The_Resistance_(game)} for a more detailed description of the game.} The game was implemented in several rounds and lasted about 30 minutes. For the purpose of our analysis, it is important to remark that all participants were seated around a circle. A tablet was placed in front of each player and it recorded the direction of the player's sight every 1/3 of a second. The structural approach we offer allows us to disentangle and estimate two key forces in explaining the data: directional sight preferences of players and peer effects in gazing.   

The two determinants of visual focus of attention that we want to recover for each person are motivated by two well-known observations: First, visual designers create online platforms under the premise that people scan certain areas of the screen before others. In particular, it is believed that people spend more time looking to the left side of the screen as compared to the right side.\footnote{See, for example, \burl{https://www.nngroup.com/articles/horizontal-attention-original-research/}.} The so-called left-to-right bias has been also documented by the experimental studies of \citet{spalek2004supporting,spalek2005left}, and \citet{reutskaja2011search}.\footnote{See also \citet{maass2003directional} and reference therein.} 
Second, in social environments, it has been observed that people automatically redirect their visual attention by following others' gaze orientation, a phenomenon called gaze following.\footnote{See, for example, \citet{gallup2012visual} and \burl{https://www.nationalgeographic.com/science/article/what-are-you-looking-at-people-follow-each-others-gazes-but-without-a-tipping-point\#close.21}.}
We believe our model is particularly well-suited to separate these two forces from the observed repeated choices. 

Before presenting the main results, we describe the dataset we use and the main assumptions we invoke in the application. 

\bigskip
\noindent\textbf{Data} The dataset contains five experiments of up to eight people. For our study, we use the data coming from the two experiments that have five players. We refer to them as Experiments 1 and 2. Experiments 1 and 2 contain 4248 and 7875 observations, respectively. Although the frequency of sight measure seems to be rather high (3 measurements per second), 19 and 26 percent of the observations in Experiments 1 and 2, respectively, have more than one player changing the direction of sight between consecutive measurements. Thus, we treat the datasets generated by the two experiments as discrete time data (i.e., Dataset 2 in Section~\ref{MI}). 
\par
\bigskip
\noindent\textbf{Choice sets} To capture possible preferences for direction in visual focus of attention, we aggregate the data to three options: left, right, and tablet.\footnote{This aggregation has several other advantages. For instance, the induced model has more players than alternatives. It also reduces the cardinality of the outcome space and the possible number of preference orders ($3^5=243$ vs. $5^5=3125$).} (The original dataset has five options as each person can look at one of her four opponents or her tablet.)  As a result, every measurement indicates the direction of sight of every player at the moment of recording (e.g., Player 1 looks to the left).  Formally, from the initial dataset we construct a new dataset on the joint configuration of choices $\mathbf{y}=(y_a)_{a\in\mathcal{A}}\in {\mathcal{Y}}^A=\times_{a\in\mathcal{A}}\{t_a,l_a,r_a\}$, where: $y_a=t_a$ if, at the moment of recording, Player $a$ looks at her own tablet; $y_a=l_a$ if Player $a$ looks to her left; and $y_a=r_a$ if Player $a$ looks to her right.
Players were seated clockwise. Thus, the left side for Players $a=1,2,3$, is defined as $t_a=\{a+1,a+2\}$ and for Players 4 and 5 it is $t_4=\{5,1\}$ and $t_5=\{1,2\}$, respectively. Similarly, the right side is defined as $r_a=\{a-1,a-2\}$ for Players $a=3,4,5$, and as $r_2=\{1,5\}$ and $r_1=\{5,4\}$ for Players 2 and 1, respectively. Note that the choice sets differ across players.\footnote{For instance, if Players 1 and 4 look to the left, they look at different sets of players: Player 1 looks at either Player 2 or Player 3, Player 4, however, looks at either Player 5 or Player 1.}
This feature can be incorporated in our model, as none of our results crucially depend on it. 
\par
\bigskip
\noindent\textbf{Consideration probabilities and choices of friends} We assume that consideration probabilities on the direction of sight (i.e., left, right, or tablet) vary with the number of peers that look in the same direction. That is, each player is more likely to consider looking at a particular area if other players are doing so. For each Player $a$ we say Player $a'$ is looking in the direction $y_a$ if $y_a\cap y_{a'}\neq \emptyset$. For example, suppose that (at the moment of deciding where to look at) Player 1 observes that $y_2=t_2$, $y_3=r_3$, $y_4=l_4$, and $y_5=l_5$. That is, $3$ players are looking in the direction of Player 1 (Players $3$, $4$, and $5$), $3$ players are looking in the direction of Player 2 (Players $2$, $3$, and $5$), only Player 4 is looking in the direction of Player $5$, and no one is looking in the direction of Players 3 and 4. As a result, we state that 3 people are looking to the ``left'' side of Player 1 ($l_1\cap y_{a'}\neq\emptyset$ only for $a'=2,3,5$), 1 person is looking to the ``right'' side of Player 1 ($r_1\cap y_{a'}\neq\emptyset$ only for $a'=4$), and 3 people are looking at Player 1 ($t_1\cap y_{a'}\neq\emptyset$ for $a'=3,4,5$). In this situation, the probability that Player 1 considers left is $\operatorname{Q}_1\left( l_1 \mid 3\right)$, right is $\operatorname{Q}_1\left( r_1 \mid 1\right)$, and tablet is $\operatorname{Q}_1\left( t_1 \mid 3\right)$.
\par
\bigskip
\noindent\textbf{Network structure} Given the small number of players and the fact that participants are playing a party game with monetary prizes, we think it is natural to assume the network is complete. That is, we assume each player is connected to all other players in the group. 
\par
\bigskip
\noindent\textbf{Preferences and the default option} Since, in this setting, there is no reason for us to treat one of the directions as the default option (i.e., always considered and the least preferred), we use the specification of the model with no default option presented in Section~\ref{default} for the estimates. Also, we do not impose any restrictions on preferences neither for each person nor across different people. Thus, in total, we have 3 possible preference orders for each player and $3^5=243$ possible combinations of preference orders for the five players.
\par
\bigskip
Before presenting the main estimates, let us display some raw data description. Using the data from the experiments we first construct marginal shares of each option for every player in Experiments 1 and 2. This information is displayed in Tables~\ref{table: marg exper1} and~\ref{table: marg exper2} below. 

\begin{table}[h]
\centering
\begin{threeparttable}
\centering
\caption{Marginal Shares in the Experiment 1 (\%)}\label{table: marg exper1}
\begin{tabular}{cccccc}
\hline
\hline
                   & Player 1 & Player 2 & Player 3 & Player 4 & Player 5\\ 
\hline
Left               & 24.58    & 50.66    & 56.57    & 78.32    & 55.25   \\
Tablet             & 6.97     & 2.71     & 0.05     & 2.68     & 0.94    \\
Right              & 68.45    & 46.63    & 43.38    & 19       & 43.81    \\
\hline
\end{tabular}
\vspace{1ex}
\begin{tablenotes}
\item {\footnotesize Notes: The sample size is 4248.}
\end{tablenotes}
\end{threeparttable}
\end{table}
Looking at Table~\ref{table: marg exper1} we see that, in Experiment 1, all the players spend very little time looking at the tablet. In addition, all players but Player 1 look to the left more often than to the right. Table~\ref{table: marg exper2} displays rather different results for Experiment 2. In particular, the second experiment has more balanced shares. Also, looking at own tablet is not longer the least selected option for all the players. For example, Player 2 looks to her tablet more often than to the left or to the right. In addition, three out of the five players look to the right more often than to the left. Based on these initial results, one may expect considerable differences in the two experiments regarding preferences for directional sight. Interestingly, we get that this is not the case.  

\begin{table}[h]
\centering
\begin{threeparttable}
\centering
\caption{Marginal Shares in the Experiment 2 (\%)}\label{table: marg exper2}
\begin{tabular}{cccccc}
\hline
\hline
                   & Player 1 & Player 2 & Player 3 & Player 4 & Player 5\\ 
\hline
Left               & 39.7     & 34.9     & 30.08    & 28.98    & 50.16   \\
Tablet             & 20.39    & 37.82    & 4.88     & 14.59    & 16.22   \\
Right              & 39.91    & 27.28    & 65.04    & 56.43    & 33.62   \\
\hline
\end{tabular}
\vspace{1ex}
\begin{tablenotes}
\item {\footnotesize Notes: The sample size is 7875.}
\end{tablenotes}
\end{threeparttable}
\end{table}
We estimate a few specifications of the model. We present two of them next and display the other ones in Appendix~\ref{app: diff spec}. In Model I(a), we assume there is no heterogeneity in consideration probabilities across options and players, i.e., $\operatorname{Q}(\cdot)=\operatorname{Q}_{a}\left( v \mid \cdot\right)$ for all $a$ and $v$. Model I(b) adds heterogeneity in consideration probabilities across players. All the models we estimate allow for unrestricted heterogeneity in preferences regarding directional visual sight.
\par
Figure~\ref{fig: Q short} shows the estimates for the consideration probabilities of Model I(a) for the two experiments. Note that although monotonicity of $\operatorname{Q}$ in the number of friends was not imposed in the estimation, the estimated probabilities are indeed monotone in Experiment 2 and are very close to be monotone in Experiment 1. The estimated preferences for directional sight coincide for all players in the two experiments. Specifically, all players prefer looking to the left, then to tablet, and then to the right. That is, we obtain that, for each $a\in\mathcal{A}$, 
\begin{equation*}
 l_a\succ_a t_a\succ_a r_a.
\end{equation*}
Thus, in the first specification of our model, all the players (in Experiments 1 and 2) show the left-to-right bias.

\begin{figure}
\centering
\begin{subfigure}{.5\textwidth}
  \centering
  \begin{tikzpicture}
    \begin{axis}[
    xmin=-0.1, xmax=4.1,
    ymin=-0.1, ymax=1.1,
    axis lines = left,
    xlabel = Number of friends,
    ylabel = Consideration probability,
    legend entries={$\operatorname{Q}$},
    legend style={at={(0.1,0.8)},anchor=west},
    ]
    \addplot[smooth,mark=*] coordinates {(0,0.032) (1,0.22) (2,0.352) (3,0.414) (4,0.495)} ;
    \end{axis}
    \end{tikzpicture}
    \caption{Experiment 1}
    \label{fig: Q short1}
\end{subfigure}%
\begin{subfigure}{.5\textwidth}
  \centering
  \begin{tikzpicture}
    \begin{axis}[
    xmin=-0.1, xmax=4.1,
    ymin=-0.1, ymax=1.1,
    axis lines = left,
    xlabel = Number of friends,
    ylabel = Consideration probability,
    legend entries={$\operatorname{Q}$},
    legend style={at={(0.1,0.8)},anchor=west},
    ]
    \addplot[smooth,mark=*] coordinates {(0,0.292) (1,0.317) (2,0.347) (3,0.385) (4,0.492)};
    \end{axis}
    \end{tikzpicture}
  \caption{Experiment 2}
  \label{fig: Q short2}
\end{subfigure}
\caption{Consideration probability as a function of the number of friends looking in the same direction. Model I(a). }
\label{fig: Q short}
\end{figure}
\par 

Figure~\ref{fig: Q long} shows the estimates for the consideration probabilities for Model I(b) in Experiments 1 and 2. Similar to the case with homogeneity, many of the consideration probabilities for each direction of sight are indeed increasing in the number of players looking at that direction. The estimated preference orders coincide in Models I(a) and (b), for the two experiments. That is, here again, all the players prefer looking to the left, then to tablet, and then to the right.

To sum up, despite considerable differences in raw marginal shares (see Tables~\ref{table: marg exper1} and~\ref{table: marg exper2}), the preferences for directional sight of all players in the two experiments coincide. Specifically, all players prefer to look from left to right after controlling for unobserved consideration sets that are determined by visual focus of attention of others. As a robustness check, in order to address possible concerns with the dynamic of players' interactions in different stages of the game, we estimated the previous two models using the last half of observations and the middle half of observations (i.e., we did not use the first and the last quarter of observations). In both cases the estimated consideration probabilities are similar to the ones estimated using the whole sample, and the estimated preferences of all players are again $l_a\succ_a t_a\succ_a r_a$ (see Appendix~\ref{app: different parts} for further details). As we mentioned earlier, this type of preferences are consistent with the left-to-right bias invoked by online platform designers that has been also documented by experimental work. 
\begin{figure}
\centering
\begin{subfigure}{.5\textwidth}
  \centering
  \begin{tikzpicture}
    \begin{axis}[
    xmin=-0.1, xmax=4.1,
    ymin=-0.1, ymax=1.1,
    axis lines = left,
    xlabel = Number of friends,
    ylabel = Consideration probability,
    legend entries={$\operatorname{Q}_1$, $\operatorname{Q}_2$, $\operatorname{Q}_3$, $\operatorname{Q}_4$, $\operatorname{Q}_5$},
    legend style={at={(0.1,0.8)},anchor=west},
    ]
    \addplot +[smooth] coordinates {(0,0.0) (1,0.144) (2,0.318) (3,0.273) (4,0.416)};
    \addplot +[smooth] coordinates {(0,0.059) (1,0.207) (2,0.223) (3,0.567) (4,0.772)};
    \addplot +[smooth] coordinates {(0,0.003) (1,0.196) (2,0.462) (3,0.502) (4,0.657)};
    \addplot +[smooth] coordinates {(0,0.103) (1,0.422) (2,0.624) (3,0.562) (4,0.642)};
    \addplot +[smooth] coordinates {(0,0.3) (1,0.349) (2,0.338) (3,0.324) (4,0.252)};
    \end{axis}
    \end{tikzpicture}
    \caption{Experiment 1}
    \label{fig: Q long1}
\end{subfigure}%
\begin{subfigure}{.5\textwidth}
  \centering
  \begin{tikzpicture}
    \begin{axis}[
    xmin=-0.1, xmax=4.1,
    ymin=-0.1, ymax=1.1,
    axis lines = left,
    xlabel = Number of friends,
    ylabel = Consideration probability,
    legend entries={$\operatorname{Q}_1$, $\operatorname{Q}_2$, $\operatorname{Q}_3$, $\operatorname{Q}_4$, $\operatorname{Q}_5$},
    legend style={at={(0.1,0.8)},anchor=west},
    ]
    \addplot +[smooth] coordinates {(0,0.381) (1,0.404) (2,0.38) (3,0.377) (4,0.432)};
    \addplot +[smooth] coordinates {(0,0.473) (1,0.432) (2,0.414) (3,0.418) (4,0.565)};
    \addplot +[smooth] coordinates {(0,0.15) (1,0.177) (2,0.222) (3,0.307) (4,0.464)};
    \addplot +[smooth] coordinates {(0,0.24) (1,0.262) (2,0.318) (3,0.405) (4,0.563)};
    \addplot +[smooth] coordinates {(0,0.391) (1,0.36) (2,0.397) (3,0.422) (4,0.519)};
    \end{axis}
    \end{tikzpicture}
    \caption{Experiment 2}
    \label{fig: Q long2}
\end{subfigure}
\caption{Consideration probabilities for different players as functions of the number of friends looking in the same direction. Model I(b).}
\label{fig: Q long}
\end{figure}

\section{Final Remarks\label{C}}

\noindent This paper adds peer effects to the consideration set models. It does so by combining the dynamic model of social interactions of \citet{blume1993statistical, blume1995statistical} with the (single-agent) model of random consideration sets of \citet{manski1977structure} and \citet{manzini2014stochastic}. The model we build differs from most of the social interaction models in that the choices of friends do not affect preferences but the subset of options that people end up considering. In our model, the network structure affects the nature and strength of the mistakes that people make. Thus, we can think of our model as a model of peer effects in mistakes.

From an applied perspective, changes in the choices of friends induce stochastic variation of the considerations sets. We show that this variation can be used to recover the main parts of the model. On top of nonparametrically recovering the preference ranking of each person and the attention mechanism, we identify the set of connections or edges between the people in the network. The identification strategy allows unrestricted heterogeneity across people. 

We propose a consistent estimator of model parameters and apply it to an experimental dataset. The structural approach we offer allows us to identify and estimate two main determinants on people visual focus of attention: directional sight preferences and peer effects in gazing. Our results are consistent with the existing literature. We believe our approach to peer effects in consideration sets could be incorporated in various empirical studies of technology adoption and diffusion. For instance, \citet{aral2009distinguishing} develop a dynamic matched sample estimation framework to distinguish influence and homophily effects in the day-by-day adoption of a mobile service application (i.e., Yahoo! Go). A similar dynamic matched sample estimation framework could be used to recover preferences over adoption and the influence of neighbor adoption rates on likelihood of considering the possibility of incorporating the mobile service application.  

\phantomsection\addcontentsline{toc}{section}{\refname}\bibliography{references}
\appendix
\section{Proofs\label{P}}

\noindent \textbf{Proof of Proposition~\ref{P1}: }For an irreducible, finite-state, continuous Markov chain the steady-state $\mu$ exists and it is unique. Thus, we only need to prove that A1 implies that the Markov chain induced by our model is irreducible. First note that, under A1, for each $a\in \mathcal{A}$, $v\in \mathcal{Y}$, and $\mathbf{y}\in \overline{\mathcal{Y}}^{A}$, we have that 
\begin{equation*}
1>\operatorname{P}_{a}\left( v \mid \mathbf{y}\right) =\operatorname{Q}_{a}\left( v \mid y_{a},\operatorname{N}_{a}^{v}\left( \mathbf{y}\right) \right)
\prod\nolimits_{v'\in \mathcal{Y},v'\succ_{a}v}\left( 1-\operatorname{Q}_{a}\left( v' \mid y_{a},\operatorname{N}_{a}^{v'}\left( 
\mathbf{y}\right) \right) \right) >0.
\end{equation*}
To show irreducibility, let $\mathbf{y}$ and $\mathbf{y}'$ be two different choice configurations. It follows from expression (\ref{T}) that we can go from one configuration to the other one in less than $A$ steps with positive probability.{\normalsize \hfill }$\blacksquare $\bigskip

\noindent \textbf{Proof of Proposition~\ref{ID1}:} By A1, $\operatorname{P}_{a}\left( \cdot \mid \mathbf{y}\right)$ has full support for each $\mathbf{y}\in\overline{\mathcal{Y}}^{A}$. By A2 and A3, $\operatorname{P}_{a}\left( v \mid \mathbf{y}\right)$ is strictly decreasing in $\operatorname{N}_{a}^{v'}\left( \mathbf{y}\right)$ for each $v'\succ_{a}v$. Thus, we can recover $\mathcal{N}_{a}$. Since this
is true for each $a\in \mathcal{A}$, we can get $\Gamma =\left( \mathcal{A},e\right)$. Also, from variation in $\operatorname{N}_{a}^{v'}\left(\mathbf{y}\right)$ for each $v'\neq v$, we can recover Person $a$'s upper level set that corresponds to option $v$. That is, 
\begin{equation*}
\left\{ v'\in \mathcal{Y}:v'\succ_{a}v\right\} 
\end{equation*}
By repeating the exercise with each alternative, we can recover $\succ_{a}$. Finally, suppose that $y_{a}^*$ is the most preferred alternative for Person $a$. Then, 
\begin{equation*}
\operatorname{P}_{a}\left( y_{a}^{\ast } \mid \mathbf{y}\right) =\operatorname{Q}_{a}\left( y_{a}^{\ast } \mid y_{a},\operatorname{N}_{a}^{y_{a}^*}\left( \mathbf{y}\right) \right) .
\end{equation*}%
It follows that we can recover $\operatorname{Q}_{a}\left( y_{a}^* \mid y_{a},\operatorname{N}_{a}^{y_{a}^*}\left( \mathbf{y}\right) \right)$ for each $\mathbf{y}\in \overline{\mathcal{Y}}^{A}$. By proceeding in descending preference ordering we can then recover $\operatorname{Q}_{a}\left( v \mid y_{a},\operatorname{N}_{a}^{v}\left( \mathbf{y}\right) \right)$ for all $v\in \mathcal{Y}$ (and each $\mathbf{y}\in\overline{\mathcal{Y}}^{A}$).{\normalsize\hfill }$\blacksquare $

\bigskip

\noindent \textbf{Proof of Proposition~\ref{ID2}:} Since $\lim\nolimits_{\Delta \rightarrow 0}\mathcal{P}\left( \Delta \right) =\mathcal{M}$, we can recover transition rate matrix from the data. Recall that 
\begin{equation*}
\operatorname{m}\left( \mathbf{y}'\mid \mathbf{y}\right) =\left\{ 
\begin{array}{lcc}
0 & \text{if} & \sum\nolimits_{a\in \mathcal{A}}\mathds{1}\left( y_{a}'\neq
y_{a}\right) >1 \\ 
\sum\nolimits_{a\in \mathcal{A}}\lambda_{a}\operatorname{P}_{a}\left( y_{a}' \mid \mathbf{y}\right) \mathds{1}\left( y_{a}'\neq y_{a}\right) & 
\text{if} & \sum\nolimits_{a\in \mathcal{A}}\mathds{1}\left( y_{a}'\neq
y_{a}\right) =1
\end{array}
\right. .
\end{equation*}
Thus, $\lambda_{a}\operatorname{P}_{a}\left( y_{a}' \mid \mathbf{y}\right) =\operatorname{m}\left( y_{a}',\mathbf{y}_{-a}\mid \mathbf{y}\right)$. It follows that we can recover $\lambda_{a}\operatorname{P}_{a}\left( v \mid \mathbf{y}\right)$ for each $v\in\overline{\mathcal{Y}}$, $\mathbf{y}\in\overline{\mathcal{Y}}^{A}$, and $a\in \mathcal{A}$. Note that, for each $\mathbf{y}\in \overline{\mathcal{Y}}^{A}$, 
\begin{equation*}
\sum\nolimits_{v\in \overline{\mathcal{Y}}}\lambda_{a}\operatorname{P}_{a}\left( v \mid \mathbf{y}\right) =\lambda_{a}\sum\nolimits_{v\in \overline{\mathcal{Y}}}\operatorname{P}_{a}\left( v \mid \mathbf{y}\right) =\lambda_{a}.
\end{equation*}
Then we can also recover $\lambda_{a}$ for each $a\in \mathcal{A}$.{\normalsize\hfill}$\blacksquare$\bigskip

\noindent \textbf{Proof of Proposition~\ref{ID3}:} This proof builds on Theorem 1 of \citet{blevins2017identifying} and Theorem 3 of \citet{blevins2018identification}. For the present case, it follows from these two theorems, that the transition rate matrix  $\mathcal{M}$ is generically identified if, in addition to the conditions in Proposition~\ref{ID3}, we have that 
\begin{equation*}
\left( Y+1\right) ^{A}-AY-1\geq \frac{1}{2}.
\end{equation*}
This condition is always satisfied if $A>1$. Thus, identification of $\mathcal{M}$ follows because, by A2, $A\geq 2$. We can then uniquely recover $\left(\operatorname{P}_{a}\right)_{a\in \mathcal{A}}$ from $\mathcal{M}$ as in the proof of Proposition~\ref{ID2}{\normalsize \hfill }$\blacksquare $\bigskip

\noindent \textbf{Proof of Proposition~\ref{IDE1}: }Note that expression (\ref{Gy}) can be rewritten as follows 
\begin{eqnarray*}
&&\operatorname{P}_{a}\left( v \mid \mathbf{y}\right) =\sum\nolimits_{\mathcal{C\subseteq Y}}\operatorname{R}_{a}\left( v \mid \mathcal{C}\right)
\prod\nolimits_{v'\in \mathcal{C}}\operatorname{Q}_{a}\left( v' \mid y_{a},\operatorname{N}_{a}^{v'}\left( \mathbf{y}\right)\right) \prod\nolimits_{v''\notin \mathcal{C}}\left( 1-\operatorname{Q}_{a}\left( v'' \mid y_{a},\operatorname{N}_{a}^{v''}\left( \mathbf{y}\right) \right) \right) = \\
&&\operatorname{Q}_{a}\left( v \mid y_{a},\operatorname{N}_{a}^{v}\left( \mathbf{y}\right) \right) \sum\nolimits_{\mathcal{C\subseteq Y},v\in \mathcal{C}}\operatorname{R}_{a}\left( v \mid \mathcal{C}\right) \prod\nolimits_{v'\in\mathcal{C\setminus }\left\{ v\right\} }\operatorname{Q}_{a}\left( v' \mid y_{a},\operatorname{N}_{a}^{v'}\left( \mathbf{y}\right)\right) \prod\nolimits_{v''\notin \mathcal{C}}\left( 1-\operatorname{Q}_{a}\left( v'' \mid y_{a},\operatorname{N}_{a}^{v''}\left( \mathbf{y}\right) \right) \right) .
\end{eqnarray*}%
Thus, by A2 and A3, we can state whether $a'\in \mathcal{N}_{a}$ by checking whether $\operatorname{P}_{a}\left( v \mid y_{1}=o,\dots,y_{A}=o\right)$ moves up when we change $y_{a'}$ from $o$ to $v$ for some $v$ in $\mathcal{Y}$. It follows that the network structure is identified.

Let $\mathbf{y}$ be such that $\operatorname{N}_{a}^{v}\left( \mathbf{y}\right) =0$ and let us assume that at least one person (different from $a$) in $\mathbf{y}$
selected the default option (i.e., there is at least one $y_{a'}=o$ with $a'\neq a$). Let $\mathbf{y}'$ be such that 
\begin{equation*}
\operatorname{N}_{a}^{v'}\left( \mathbf{y}\right) =\operatorname{N}_{a}^{v'}\left( \mathbf{y}'\right) \text{ for all }v'\neq v\text{ and }\operatorname{N}_{a}^{v}\left( \mathbf{y}\right) =1.
\end{equation*}
Note that 
\begin{equation*}
\operatorname{P}_{a}\left( v \mid \mathbf{y}'\right) /\operatorname{P}
_{a}\left( v \mid \mathbf{y}\right) =\operatorname{Q}_{a}\left( v \mid y_{a},1\right) /\operatorname{Q}_{a}\left( v \mid y_{a},0\right) .
\end{equation*}
Also 
\begin{equation*}
\operatorname{P}_{a}\left( o \mid \mathbf{y}'\right) /\operatorname{P}
_{a}\left( o \mid \mathbf{y}\right) =\left( 1-\operatorname{Q}_{a}\left( v 
 \mid y_{a},1\right) \right) /\left( 1-\operatorname{Q}_{a}\left( v 
 \mid y_{a},0\right) \right) .
\end{equation*}
Thus, by A3, $\operatorname{Q}_{a}\left( v \mid y_{a},0\right)$ and $\operatorname{Q}_{a}\left( v \mid y_{a},1\right)$ can be recovered from the
data. By implementing a similar procedure for different values of $\operatorname{N}_{a}^{v}\left( \mathbf{y}\right)$ we can recover $\operatorname{Q}_{a}$ for each $y_{a}$.
Finally, since this is true for any arbitrary $a$ and $y_{a}$, then we can
recover the attention mechanism $\left( \operatorname{Q}_{a}\right)_{a\in \mathcal{A}}$.

We finally show that $\operatorname{R}_{a}$ is identified if and only if (in addition to A1-A3) we have that $\left\vert \mathcal{N}_{a}\right\vert \geq Y-1$. We will present the idea for $v=1$, agent $a$, and $y_{a}$. (The proof immediately extends to other agents and alternatives.) We want to recover $\operatorname{R}_{a}\left( 1 \mid \mathcal{C}\right)$ for all $\mathcal{C}$. To simplify the exposition we will write 
\begin{eqnarray*}
\left\vert \mathcal{N}_{a}\right\vert &=&N \\
\operatorname{Q}_{a}\left( v \mid y_{a},m\right) &=&\operatorname{Q}^{1}\left( v \mid m\right) \\
1-\operatorname{Q}_{a}\left( v \mid y_{a},m\right) &=&\operatorname{Q}^{0}\left( v \mid m\right)
\end{eqnarray*}
We have a set of equations indexed by $\mathbf{y}$ 
\begin{equation*}
\operatorname{P}_{a}\left( 1 \mid \mathbf{y}\right) /\operatorname{Q}_{a}\left( 1 \mid y_{a},\operatorname{N}_{a}^{1}\left( \mathbf{y}\right) \right)
=\sum\nolimits_{\mathcal{C\subseteq Y},v\in \mathcal{C}}\operatorname{R}_{a}\left( v \mid \mathcal{C}\right) \prod\nolimits_{k\in \mathcal{C\setminus }\left\{v\right\} }\operatorname{Q}^{1}\left( k \mid \operatorname{N}_{a}^{k}\left( \mathbf{y}\right) \right) \prod\nolimits_{k\notin \mathcal{C\setminus }\left\{v\right\} }\operatorname{Q}^{0}\left( k \mid \operatorname{N}_{a}^{k}\left( \mathbf{y}\right) \right) .
\end{equation*}%
To present the ideas more clear let $A(N,Y)$ be the matrix of coefficients in front of the $\operatorname{R}_{a}^{'}$s. The above system of equations has a unique solution if and only if $A(N,Y)$ has full column rank. The column of $A(N,Y)$ that corresponds to any given $\mathcal{C\subseteq Y}$ consists of
the elements of the following form 
\begin{equation*}
\prod\nolimits_{k\in \mathcal{Y}}\operatorname{Q}^{\mathds{1}\left( k\in \mathcal{C}\right)
}\left( k \mid \operatorname{N}^{k}\right)
\end{equation*}%
where $\operatorname{N}^{k}\in \left\{ 0,1,\dots,N\right\}$ and $\sum\nolimits_{k}\operatorname{N}^{k}\leq N$. The last claim in the proposition follows from the next lemma.\bigskip

\noindent \textbf{Lemma 1}: Assume that A1-A3 hold. For all $Y\geq 2$ and $N\geq 1$ 
\begin{equation*}
N\geq Y-1\Longleftrightarrow A(N,Y)\text{ has full column rank.}
\end{equation*}

\noindent \textbf{Proof.}\bigskip

\noindent \textbf{Step 1. }To illustrate how the idea works, note that $%
A(1,2)$ and $A(1,3)$ can be written as follows 
\begin{equation*}
A(1,2)=\left( 
\begin{array}{cc}
\operatorname{Q}^{0}\left( 2 \mid 0\right) & \operatorname{Q}^{1}\left( 2 
 \mid 0\right) \\ 
\operatorname{Q}^{0}\left( 2 \mid 1\right) & \operatorname{Q}^{1}\left( 2 
 \mid 1\right)%
\end{array}
\right)
\end{equation*}
and 
\begin{eqnarray*}
A(1,3) &=&\left( 
\begin{array}{cccc}
\operatorname{Q}^{0}\left( 3 \mid 0\right) \operatorname{Q}^{0}\left( 2 \mid 0\right) & \operatorname{Q}^{0}\left( 3 \mid 0\right) \operatorname{Q}^{1}\left( 2 \mid 0\right) & \operatorname{Q}^{1}\left( 3 \mid 0\right) \operatorname{Q}^{0}\left( 2 \mid 0\right) & \operatorname{Q}^{1}\left( 3 \mid 0\right) \operatorname{Q}^{1}\left( 2 \mid 0\right) \\ 
\operatorname{Q}^{0}\left( 3 \mid 0\right) \operatorname{Q}^{0}\left( 2 \mid 1\right) & \operatorname{Q}^{0}\left( 3 \mid 0\right) \operatorname{Q}^{1}\left( 2 \mid 1\right) & \operatorname{Q}^{1}\left( 3 \mid 0\right) \operatorname{Q}^{0}\left( 2 \mid 1\right) & \operatorname{Q}^{1}\left( 3 \mid 0\right) \operatorname{Q}^{1}\left( 2 \mid 1\right) \\ 
\operatorname{Q}^{0}\left( 3 \mid 1\right) \operatorname{Q}^{0}\left( 2 \mid 0\right) & \operatorname{Q}^{0}\left( 3 \mid 1\right) \operatorname{Q}^{1}\left( 2 \mid 0\right) & \operatorname{Q}^{1}\left( 3 \mid 1\right) \operatorname{Q}^{0}\left( 2 \mid 0\right) & \operatorname{Q}^{1}\left( 3 \mid 1\right) \operatorname{Q}^{1}\left( 2 \mid 0\right)
\end{array}
\right) \\
&=&\left( 
\begin{array}{cc}
\operatorname{Q}^{0}\left( 3 \mid 0\right) A(1,2) & \operatorname{Q}^{1}\left( 3 \mid 0\right) A(1,2) \\ 
\operatorname{Q}^{0}\left( 3 \mid 1\right) A(0,2) & \operatorname{Q}^{1}\left( 3 \mid 1\right) A(0,2)
\end{array}
\right)
\end{eqnarray*}
where $A(0,2)=\left( 
\begin{array}{cc}
\operatorname{Q}^{0}\left( 2 \mid 0\right) & \operatorname{Q}^{1}\left( 2 
 \mid 0\right)%
\end{array}
\right)$.

Similarly, the matrix $A(N,Y+1)$ can be written as follows 
\begin{equation*}
A(N,Y+1)=\left( 
\begin{array}{cc}
\operatorname{Q}^{0}\left( Y+1 \mid 0\right) A(N,Y) & \operatorname{Q}
^{1}\left( Y+1 \mid 0\right) A(N,Y) \\ 
\operatorname{Q}^{0}\left( Y+1 \mid 1\right) A(N-1,Y) & \operatorname{Q}
^{1}\left( Y+1 \mid 1\right) A(N-1,Y) \\ 
\operatorname{Q}^{0}\left( Y+1 \mid 2\right) A(N-2,Y) & \operatorname{Q}
^{1}\left( Y+1 \mid 2\right) A(N-2,Y) \\ 
\dots & \dots \\ 
\operatorname{Q}^{0}\left( Y+1 \mid N\right) A(0,Y) & \operatorname{Q}^{1}\left(
Y+1  \mid N\right) A(0,Y)%
\end{array}
\right) .
\end{equation*}
Note that $A(K,Y)$ is a submatrix of $A(K+1,Y)$ for all $K$ (with the same number of columns). Thus, it is clear that $A(N,Y+1)$ has full column rank only if $A(N,Y)$ and $A(N-1,Y)$ have both full column rank, which is the same as to say $A(N-1,Y)$ has full column rank. We next show that under A3, if $A(N-1,Y)$ has full column rank, then $A(N,Y+1)$ has full column rank too. To this end, let $M$ be a matrix obtained deleting rows from $A(N-1,Y)$ in such a way that $\det \left( M\right) >0$. Then, by A3, we have 
\begin{equation*}
\det \left( 
\begin{array}{cc}
\operatorname{Q}^{0}\left( Y+1 \mid 0\right) M & \operatorname{Q}^{1}\left(
Y+1 \mid 0\right) M \\ 
\operatorname{Q}^{0}\left( Y+1 \mid 1\right) M & \operatorname{Q}^{1}\left(
Y+1 \mid 1\right) M%
\end{array}
\right) =\left( \operatorname{Q}^{1}\left( Y+1 \mid 1\right) -\operatorname{Q}^{1}\left( Y+1 \mid 0\right) \right) ^{2^{Y-1}}\det\left( M\right) ^{2}>0.
\end{equation*}
In summary, we have that 
\begin{equation*}
A(N,Y+1)\text{ has full column rank }\Longleftrightarrow A(N-1,Y)\text{ has
full column rank.}
\end{equation*}

\noindent \textbf{Step 2. }Consider $(N,Y)=(1,2)$. Note that 
\begin{equation*}
A(1,2)=\left( 
\begin{array}{cc}
\operatorname{Q}^{0}\left( 2 \mid 0\right) & \operatorname{Q}^{1}\left( 2 \mid 0\right) \\ 
\operatorname{Q}^{0}\left( 2 \mid 1\right) & \operatorname{Q}^{1}\left( 2 \mid 1\right)
\end{array}
\right)
\end{equation*}
has full column rank since $\det \left( A(1,2)\right) =\operatorname{Q}^{1}\left( 2 \mid 1\right) -\operatorname{Q}^{1}\left( 2 \mid 0 \right) >0$. Also, any $A(N,2)$ with $N\geq 1$ will have full column rank because $A(1,2)$ is a submatrix of $A(N,2)$ with the same number of columns.

Finally, note that $A(1,3)$ does not have full column rank since the number of columns is higher than the number of rows. \bigskip

\noindent \textbf{Step 3. }From Step 1\textbf{\ }we got that, for all $Y\geq2$ and $N\geq 1$, 
\begin{equation*}
A(N,Y+1)\text{ has full column rank }\Longleftrightarrow A(N-1,Y)\text{ has
full column rank.}
\end{equation*}%
From Step 2, we get that $A(N,2)$ (with $N\geq 1$) has full column rank and $A(1,3)$ does not have full column rank. The claim in Lemma 1 follows by
combining these three results. {\normalsize \hfill }$\blacksquare $

\bigskip

\noindent \textbf{Proof of Proposition~\ref{IDE2}:} This proof is divided in three steps.\bigskip

\noindent \textbf{Step 1. (Identification of the Set of Connections)} Take any two different people with arbitrary designations $a_{1}$ and $a_{2}$. Note that if $a_{2}\not\in \mathcal{N}_{a_{1}}$, then $\operatorname{P}_{a_{1}}(v \mid \mathbf{y})=\operatorname{P}_{a_{1}}(v \mid \mathbf{y}')$ for any $\mathbf{y}$ and $\mathbf{y}'$ such that $y_{a}=y_{a}'$ for all $a\neq a_{2}$ and $y_{a_{1}}\neq v$. Also, let $v_{a_{1}}^*$ be the best preferred alternative of $a_{1}$. Then, by A1 and A3, for any $\mathbf{y}$ such that $y_{a_{1}}\neq v_{a_{1}}^*$ 
\begin{equation*}
\operatorname{P}_{a_{1}}\left( v_{a_{1}}^{\ast } \mid \mathbf{y}\right) =\operatorname{Q}_{a_{1}}\left( v_{a_{1}}^{\ast } \mid \mathbf{y}\right)
\end{equation*}
is constant in $y_{a_{2}}$ if and only if $a_{2}\not\in \mathcal{N}_{a_{1}}$. Altogether, $a_{2}\not\in \mathcal{N}_{a_{1}}$ if and only if $\operatorname{P}_{a_{1}}\left( v \mid \mathbf{y}\right)$ with $y_{a_{1}}\neq v$ is constant in $y_{a_{2}}$. As a result, we can identify whether $a_{2}$ is in the set of friends of $a_{1}$. Since the choice of $a_{1}$ and $a_{2}$ was arbitrary, we can identify the set of connections $\Gamma $. Note that for this result we only need $Y\geq 2$. \bigskip

\noindent \textbf{Step 2. (Identification of the Preferences)} Fix some Person $a_{1}$. We will show the result for a set of alternatives $\mathcal{Y=}\left\{ 1,2,3\right\} $ of size $3$. (The proof easily extends to the case of more alternatives. The only cost is extra notation.) For arbitrary designation of the three options $v_{1},v_{2},v_{3}$, and any $a_{2}\in \mathcal{N}_{a_{1}}$ (by A2, $\mathcal{N}_{a_{1}}\neq \emptyset $) let 
\begin{equation*}
\Delta \operatorname{P}(v_{1},v_{2},v_{3})=\operatorname{P}_{a_{1}}(v_{1} \mid (v_{3},\dots
,v_{3},v_{2},v_{3},\dots ,v_{3}))-\operatorname{P}_{a_{1}}(v_{1} \mid (v_{3},\dots
,v_{3},v_{3},v_{3},\dots ,v_{3})),
\end{equation*}%
where Person $a_{2}$ switches from $v_{2}$ to $v_{3}$. Define $\mathrm{sign}(v_{1},v_{2},v_{3})\in \{-,+,0\}$ be such that 
\begin{equation*}
\mathrm{sign}(v_{1},v_{2},v_{3})=%
\begin{cases}
- & \text{ if }\Delta \operatorname{P}(v_{1},v_{2},v_{3})<0 \\ 
+ & \text{ if }\Delta \operatorname{P}(v_{1},v_{2},v_{3})>0 \\ 
0 & \text{ if }\Delta \operatorname{P}(v_{1},v_{2},v_{3})=0
\end{cases}
.
\end{equation*}
Note that $\mathrm{sign}(\cdot )$ can be computed from the data. Different preference orders over $\mathcal{Y}$ will imply potentially different values for $\mathrm{sign}(\cdot )$. Let $\operatorname{N}_{a_{1}}$ is the the cardinality of $\mathcal{N}_{a_{1}}$. For example, if $1\succ_{a_{1}}2\succ_{a_{1}}3$, then we have that 
\begin{align*}
\Delta \operatorname{P}(1,2,3)& =\operatorname{Q}_{a_{1}}(1 \mid 3,0)-\operatorname{Q}_{a_{1}}(1 \mid 3,0)=0 \\
\Delta \operatorname{P}(1,3,2)& =\operatorname{Q}_{a_{1}}(1 \mid 2,0)-\operatorname{Q}_{a_{1}}(1 \mid 3,0)=0 \\
\Delta \operatorname{P}(3,1,2)& =\operatorname{Q}_{a_{1}}(3 \mid 2,0)(1-\operatorname{Q}_{a_{1}}(1 \mid 2,1))(1-\operatorname{Q}_{a_{1}}(2 \mid 2,\operatorname{N}_{a_{1}}-1)) \\
& -\operatorname{Q}_{a_{1}}(3 \mid 2,0)(1-\operatorname{Q}_{a_{1}}(1 \mid 2,0))(1-\operatorname{Q}_{a_{1}}(2 \mid 2,\operatorname{N}_{a_{1}})) \\
\Delta \operatorname{P}(3,2,1)& =\operatorname{Q}_{a_{1}}(3 \mid 1,0)(1-\operatorname{Q}_{a_{1}}(1 \mid 1,\operatorname{N}_{a_{1}}-1))(1-\operatorname{Q}_{a_{1}}(2 \mid 1,1)) \\
& -\operatorname{Q}_{a_{1}}(3 \mid 1,0)(1-\operatorname{Q}_{a_{1}}(1 \mid 1,\operatorname{N}_{a_{1}}))(1-\operatorname{Q}_{a_{1}}(2 \mid 1,0)),
\end{align*}%
Note that for $\Delta \operatorname{P}(3,1,2)$ and $\Delta \operatorname{P}(3,2,1)$ our monotonicity restriction is not informative enough: $\Delta \operatorname{P}(3,1,2)$ and $\Delta \operatorname{P}(3,2,1)$ can be positive, negative, or equal to zero. The following table displays the values $\mathrm{sign}(\cdot )$ takes depending on the underlying preference order $\succ_{a_{1}}$ for all distinct $v_{1}$, $v_{2}$, and $v_{3}$. If the sign of $\Delta\operatorname{P}(v_{1},v_{2},v_{3})$ is not uniquely determined by a strict preference, then we write $\sim$. 
\begin{equation*}
\begin{tabular}{c|c|c|c|c|c|c|}
order/$(v_{1},v_{2},v_{3})$ & $(1,2,3)$ & $(1,3,2)$ & $(2,1,3)$ & $(2,3,1)$
& $(3,1,2)$ & $(3,2,1)$ \\ \hline
$1\succ_{a_{1}}2\succ_{a_{1}}3$ & $0$ & $0$ & $-$ & $+$ & $\sim $ & $\sim $
\\ 
$1\succ_{a_{1}}3\succ_{a_{1}}2$ & $0$ & $0$ & $\sim $ & $\sim $ & $-$ & $+$
\\ 
$2\succ_{a_{1}}1\succ_{a_{1}}3$ & $-$ & $+$ & $0$ & $0$ & $\sim $ & $\sim $
\\ 
$2\succ_{a_{1}}3\succ_{a_{1}}1$ & $\sim $ & $\sim $ & $0$ & $0$ & $+$ & $-$
\\ 
$3\succ_{a_{1}}1\succ_{a_{1}}2$ & $+$ & $-$ & $\sim $ & $\sim $ & $0$ & $0$
\\ 
$3\succ_{a_{1}}2\succ_{a_{1}}1$ & $\sim $ & $\sim $ & $+$ & $-$ & $0$ & $0$
\\ \hline
\end{tabular}%
\end{equation*}%
Note that we can always distinguish $1\succ_{a_{1}}2\succ_{a_{1}}3$ from say $3\succ_{a_{1}}2\succ_{a_{1}}1$ since $\mathrm{sign}(2,1,3)$ is determined for both preference orders and gives different predictions. The only pairs of preference orders that may be observationally equivalent are the following three pairs: (i) $1\succ_{a_{1}}2\succ_{a_{1}}3$ and $1\succ_{a_{1}}3\succ_{a_{1}}2$; (ii) $2\succ_{a_{1}}1\succ_{a_{1}}3$ and $2\succ_{a_{1}}3\succ_{a_{1}}1$; and (iii) $3\succ_{a_{1}}1\succ_{a_{1}}2$ and $3\succ_{a_{1}}2\succ_{a_{1}}1$. Although we cannot uniquely identify the order, we can uniquely identify the most preferred alternative. For example, if we know that $1\succ_{a_{1}}2\succ_{a_{1}}3$ or $1\succ_{a_{1}}2\succ_{a_{1}}3$ has generated the data, then $1$ is the most preferred alternative for agent $a_{1}$.

Assume, without loss of generality, that $3$ is the most preferred alternative. Then we can identify $\operatorname{Q}_{a_{1}}(3 \mid y_{a_{1}},\cdot )$ for any $y_{a_{1}}\neq 3$ since 
\begin{equation*}
\operatorname{Q}_{a_{1}}\left(3 \mid y_{a_{1}},\operatorname{N}_{a_{1}}^{3}(\mathbf{y})\right)=\operatorname{P}_{a_{1}}(3 \mid \mathbf{y}).
\end{equation*}
Hence, for $\mathbf{y}$ with $y_{a_{1}}=1$ we have that 
\begin{equation*}
\dfrac{\operatorname{P}_{a_{1}}(2 \mid \mathbf{y})}{1-\operatorname{P}_{a_{1}}(3 \mid \mathbf{y})}=%
\begin{cases}
\operatorname{Q}_{a}(2 \mid 1,\operatorname{N}_{a_{1}}^{2}(\mathbf{y})), & \text{ if } 2\succ_{a_{1}} 1 \\ 
\operatorname{Q}_{a}(2 \mid 1,\operatorname{N}_{a_{1}}^{2}(\mathbf{y}))(1-\operatorname{Q}_{a}(1 \mid 1,\operatorname{N}_{a_{1}}^{1}(\mathbf{y}))), & \text{ if } 1\succ_{a_{1}}2.
\end{cases}
\end{equation*}
Start with a $\mathbf{y}$ such that $y_{a}=3$ for all $a\neq a_{1}$ (and $y_{a_{1}}=1$). Consider then changing the $y_{a_{2}}$ from 3 to 1 for some $a_{2}\in \mathcal{N}_{a_{1}}$. Then, by A3, $1\succ_{a_{1}}2$ if and only if $\operatorname{P}_{a_{1}}(2 \mid \mathbf{y})/\left(1-\operatorname{P}_{a_{1}}(3 \mid \mathbf{y})\right)$ strictly decreases in the data. \bigskip

\noindent \textbf{Step 3. (Identification of the Attention Mechanism)} Fix some $a_{1}$ and let $y_{a_{1}}^{\ast }$ be the most preferred alternative of Person $a_{1}$. Then we can identify $\operatorname{Q}_{a_{1}}(y_{a_{1}}^* \mid y_{a_{1}},\operatorname{N}_{a_{1}}^{y_{a_{1}}^*}(\mathbf{y}))$ for any $\mathbf{y}$ such that $y_{a_{1}}^*\neq y_{a_{1}}$ since 
\begin{equation*}
\operatorname{Q}_{a_{1}}\left(y_{a_{1}}^{\ast } \mid y_{a_{1}},\operatorname{N}_{a_{1}}^{y_{a_{1}}^*}(\mathbf{y})\right)=\operatorname{P}_{a_{1}}(y_{a_{1}}^{\ast } \mid \mathbf{y}).
\end{equation*}%
By proceeding in decreasing preference order we can recover $\operatorname{Q}_{a_{1}}(y_{a_{1}}' \mid y_{a_{1}},N_{a_{1}}^{y_{a_{1}}'}(\mathbf{y}))$ for any $y_{a_{1}}'$ and $\mathbf{y}$ such that $y_{a_{1}}'\neq y_{a_{1}}$. Moreover, we can identify 
\begin{equation*}
\prod\nolimits_{v'\neq y_{a_{1}}}\left( 1-\operatorname{Q}_{a}\left( v'%
 \mid y_{a_{1}},\operatorname{N}_{a_{1}}^{v'}(\mathbf{y})\right)
\right)
\end{equation*}

Next note that for any $\mathbf{y}$ such that $y_{a_{1}}^{*}=y_{a_{1}}$ 
\begin{equation*}
\operatorname{Q}_{a_{1}}(y_{a_{1}}^{\ast } \mid y_{a_{1}},\operatorname{N}_{a_{1}}^{y_{a_{1}}^*}(\mathbf{y}))=\dfrac{\operatorname{P}_{a_{1}}(y_{a_{1}}^{\ast } \mid \mathbf{y})-\prod\nolimits_{v'\neq y_{a_{1}}^{\ast }}\left( 1-\operatorname{Q}_{a}\left(v' \mid y_{a_{1}}^*,\operatorname{N}_{a_{1}}^{v'}(\mathbf{y})\right) \right) }{1-\prod\nolimits_{v'\neq y_{a_{1}}^*}\left( 1-\operatorname{Q}_{a}\left( v' \mid y_{a_{1}}^*,\operatorname{N}_{a_{1}}^{v'}(\mathbf{y})\right) \right) }.
\end{equation*}
Hence, we can identify $\operatorname{Q}_{a_{1}}(y_{a_{1}}^{\ast } \mid y_{a_{1}},\operatorname{N}_{a_{1}}^{y_{a_{1}}^{\ast }}(\mathbf{y}))$ for all $\mathbf{y}$. Let $y_{a_{1}}^{\ast \ast }$ be the second best alternative of Person $a_{1}$, then for any $\mathbf{y}$ such that $y_{a_{1}}^{\ast \ast }=y_{a_{1}}$ similarly to the case with $y_{a_{1}}^{\ast }$ we can identify $\operatorname{Q}_{a_{1}}(y_{a_{1}}^{\ast \ast } \mid y_{a_{1}},\operatorname{N}_{a_{1}}^{y_{a_{1}}^{\ast \ast}}(\mathbf{y}))$ since 
\begin{equation*}
\operatorname{Q}_{a_{1}}(y_{a_{1}}^{\ast \ast } \mid y_{a_{1}},N_{a_{1}}^{y_{a_{1}}^{\ast
\ast }}(\mathbf{y}))=\dfrac{\operatorname{P}_{a_{1}}(y_{a_{1}}^{\ast \ast } \mid \mathbf{y})-\prod\nolimits_{v'\neq y_{a_{1}}^{\ast \ast }}\left( 1-\operatorname{Q}_{a_{1}}\left( v' \mid y_{a_{1}}^{\ast \ast },\operatorname{N}_{a_{1}}^{v'}(\mathbf{y})\right) \right) }{1-\operatorname{Q}_{a_{1}}(y_{a_{1}}^{\ast } \mid y_{a_{1}},\operatorname{N}_{a_{1}}^{v'}(\mathbf{y}))-\prod\nolimits_{v'\neq y_{a_{1}}^{\ast \ast }}\left( 1-\operatorname{Q}_{a_{1}}\left( v' \mid y_{a_{1}}^{\ast \ast },\operatorname{N}_{a_{1}}^{v'}(\mathbf{y})\right) \right) },
\end{equation*}
and thus we recover $\operatorname{Q}_{a_{1}}(y_{a_{1}}^{\ast \ast} \mid y_{a_{1}},N_{a_{1}}^{y_{a_{1}}^{\ast \ast }}(\mathbf{y}))$ for all $\mathbf{y}$. By proceeding in decreasing preference order we can recover $\operatorname{Q}_{a_{1}}(y_{a_{1}}' \mid y_{a_{1}},N_{a_{1}}^{y_{a_{1}}'}(\mathbf{y}))$ for any $y_{a_{1}}'$ and $\mathbf{y}$. Since the choice of $a_{1}$ was arbitrary we can identify $(\operatorname{Q}_{a})_{a\in \mathcal{A}}$.

\bigskip 

\noindent \textbf{Proof of Proposition~\ref{IDE3}: }
\noindent \textbf{Step 1. (Identification of the Set of Connections)} Take any two different agents $a_1$ and $a_2$. Note that if $a_2\not\in \mathcal{N}_{a_1}$, then $\operatorname{P}_{a_1}(v \mid \mathbf{y})=\operatorname{P}_{a_1}(v \mid \mathbf{y}')$ for any $\mathbf{y}$ and $\mathbf{y}'$ such that $y_a=y'_a$ for all $a\neq a_2$ and $y_{a_1}\neq v$. Also, if $v^*_{a_1}$ is the best preferred alternative of $a_1$, then by Assumptions A1' and A3', for any $\mathbf{y}$
\[
\dfrac{\operatorname{P}_{a_1}\left( v_{a_1}^{*} \mid \mathbf{y}\right)}{\operatorname{P}_{a_1}\left(o \mid \mathbf{y}\right)} =\sum\nolimits_{\mathcal{C}\in 2^{\mathcal{Y}}:\:v_{a_1}^{*}\in\mathcal{C}}\eta_{a_1}\left( \mathcal{C} \mid \mathbf{y}\right)
\]
is constant in choices of Person $a_{2}$ if and only if $a_2\not\in \mathcal{N}_{a_1}$. Hence, $a_2\not\in \mathcal{N}_{a_1}$ if and only if $\operatorname{P}_{a_1}\left( v \mid \mathbf{y}\right)/\operatorname{P}_{a_1}\left(o \mid \mathbf{y}\right)$ is constant in $y_{a_2}$. As a result, we can identify whether $a_2$ is a friend of $a_1$. Since the choice of $a_1$ and $a_2$ was arbitrary we can identify the whole $\Gamma$. Note that for this result to hold we only need $|\mathcal{Y}|\geq 2$.
\par
\noindent \textbf{Step 2. (Identification of the Preferences)} Fix Person $a_1$ and $\mathbf{y}^*=(o,o,\dots,o)\prime$. Note that 
\[
\dfrac{\operatorname{P}_{a_1}\left( v_{a_1}^{*} \mid \mathbf{y}^*\right)}{\operatorname{P}_{a_1}\left(o \mid \mathbf{y}^*\right)} =\sum\nolimits_{\mathcal{C}\in 2^{\mathcal{Y}}:\:v_{a_1}^{*}\in\mathcal{C}}\eta_{a_1}\left( \mathcal{C} \mid \mathbf{y}^*\right)
\]
will increase if any of friends of $a_1$ switches to anything else. Let $v_{a_1}^{**}$ be the second best preferred alternative of $a_1$. Then
\[
\dfrac{\operatorname{P}_{a_1}\left( v_{a_1}^{**} \mid \mathbf{y}^*\right)}{\operatorname{P}_{a_1}\left(o \mid \mathbf{y}^*\right)} =\sum\nolimits_{\mathcal{C}\in 2^{\mathcal{Y}}:\:v_{a_1}^{*}\not\in\mathcal{C},\:v_{a_1}^{**}\in\mathcal{C}}\eta_{a_1}\left( \mathcal{C} \mid \mathbf{y}^*\right)
\]
will increase if any of friends of $a_1$ switches to anything else but $v_{a_1}^{*}$. Moreover, this probability will not change if any of friends of $a_1$ switches to $v_{a_1}^{*}$. Hence, we can identify $v_{a_1}^{*}$. Applying the above step in decreasing order, we can identify the whole preference order of Person $a_1$. Since the choice of $a_1$ was arbitrary we identify preferences of all persons. 
\par
\noindent \textbf{Step 3. (Identification of the Attention Mechanism)} Take any Person $a_{1}$ and configuration $\mathbf{y}$. Since preferences are identified, assume without loss of generality that $Y\succ_{a_{1}} Y-1\succ_{a_{1}}Y-2\succ_{a_{1}}\dots\succ_{a_{1}}2\succ_{a_{1}}1$. Note that we have the following system of $Y$ equations
\[
\dfrac{\operatorname{P}_{a_1}\left( k \mid \mathbf{y}\right)}{\operatorname{P}_{a_1}\left(o \mid \mathbf{y}\right)}=\begin{cases}
\eta_{a_{1}}(\{1\} \mid \mathbf{y}),&k=1,\\
\sum\nolimits_{\mathcal{C}\subseteq \{1,\dots,k-1\}}\eta_{a_{1}}(\mathcal{C}\cup\{k\} \mid \mathbf{y}),&k=2,\dots,Y.
\end{cases}
\]
Note that $\eta_{a_{1}}$ could have generated the data if and only if it solves the above system of equations. Since, there are $2^Y-1$ unknown parameters (recall that attention to the empty set is normalized to be 1) and $Y$ equations, and there is no single attention parameter that enters more than one equation, $\eta_{a_{1}}$ can not be identified without more restrictions. Suppose $\eta_{a_{1}}$ is multiplicative. So we only need to identify $\eta_{a_{1}}(\{k\} \mid \mathbf{y})$ for all $k\in\mathcal{Y}$. Then, first, we can identify $\eta_{a_{1}}(\{1\} \mid \mathbf{y})$ since
\[
\dfrac{\operatorname{P}_{a_1}\left( 1 \mid \mathbf{y}\right)}{\operatorname{P}_{a_1}\left(o \mid \mathbf{y}\right)}= \eta_{a_{1}}(\{1\} \mid \mathbf{y}).
\]
Next, 
\[
\dfrac{\operatorname{P}_{a_1}\left( 2 \mid \mathbf{y}\right)}{\operatorname{P}_{a_1}\left(o \mid \mathbf{y}\right)}= \eta_{a_{1}}(\{2\} \mid \mathbf{y})+\eta_{a_{1}}(\{1,2\} \mid \mathbf{y})=\eta_{a_{1}}(\{2\} \mid \mathbf{y})+\eta_{a_{1}}(\{1\} \mid \mathbf{y})\eta_{a_{1}}(\{2\} \mid \mathbf{y}).
\]
Hence, we identify $\eta_{a_{1}}(\{2\} \mid \mathbf{y})$. Repeating the above steps recursively we can identify $\eta_{a_{1}}$ since
\[
\eta_{a_{1}}(\{k\} \mid \mathbf{y})=\dfrac{\operatorname{P}_{a_1}\left( k \mid \mathbf{y}\right)}{\operatorname{P}_{a_1}\left(o \mid \mathbf{y}\right)}\cdot\dfrac{1}{\sum\nolimits_{\mathcal{C}\subseteq \{1,\dots,k-1\}}\eta_{a}(\mathcal{C} \mid \mathbf{y})}.
\]
The same argument can be applied if $\eta_{a_{1}}$ is additive. This approach can be generalized if one models the attention index of a set as a strictly increasing transformation of attention indexes of elements of that set (i.e., $\eta_{a_{1}}(\{1,2\} \mid \mathbf{y})=\phi_{\{1,2\}}(\eta_{a_{1}}(\{1\} \mid \mathbf{y}),\eta_{a_{1}}(\{2\} \mid \mathbf{y}))$, where $\phi_{1,2}$ is strictly increasing in both arguments).
\par
Suppose that $\eta_{a_{1}}$ is the same for sets of the same cardinality. Since we know $\eta_{a_{1}}(\{1\})$ from the first equation we identify attention indexes for all singleton sets including $\eta_{a_{1}}(\{2\} \mid \mathbf{y})$. Hence, using the second equation we identify $\eta_{a_{1}}(\{1,2\} \mid \mathbf{y})$ and, thus, attention indexes for all sets of cardinality $2$. Repeating the above arguments recursively we can identify $\eta_{a_{1}}$. 
\par
It is left to show that if $\eta_{a_{1}}$ is the same for sets that have the same best option, then it is also identified. Again, $\eta_{a_{1}}(\{1\})$is identified from the first equation. Since $\eta_{a_{1}}(\{2\} \mid \mathbf{y})=\eta_{a_{1}}(\{1,2\} \mid \mathbf{y})$, from the second equation we identify
\[
\eta_{a_{1}}(\{2\} \mid \mathbf{y})=\eta_{a_{1}}(\{1,2\} \mid \mathbf{y})=\dfrac{\operatorname{P}_{a_{1}}\left( 2 \mid \mathbf{y}\right)}{2\operatorname{P}_{a_{1}}\left(o \mid \mathbf{y}\right)}.
\]
Repeating the above arguments for every equation we get that
\[
\eta_{a_{1}}(\mathcal{C} \mid \mathbf{y})=\dfrac{\operatorname{P}_{a_{1}}\left( v^*_{a_1,\mathcal{C}} \mid \mathbf{y}\right)}{v^*_{a_1,\mathcal{C}}\operatorname{P}_{a_{1}}\left(o \mid \mathbf{y}\right)},
\]
where $v^*_{a_1,\mathcal{C}}$ is the best alternative in $\mathcal{C}$ according to $\succ_{a_1}$. Since the choice of $a_1$ and $\mathbf{y}$ was arbitrary, we identify $\eta_{a}$ for all $a\in\mathcal{A}$.
\par
We conclude the proof by showing that restrictions (i) and (ii) correspond to the models of consideration sets formation in \citet{manzini2014stochastic} and \citet{dardanoni2020inferring}, respectively. Equivalence between (ii) and the model in \citet{dardanoni2020inferring} is straightforward since 
\begin{equation*}
\dfrac{\eta_a(\mathcal{C} \mid \mathbf{y})}{\displaystyle\sum\nolimits_{\mathcal{B}\subseteq\mathcal{Y}}\eta_a(\mathcal{B} \mid \mathbf{y})}=\dfrac{\eta_a(\mathcal{D} \mid \mathbf{y})}{\displaystyle\sum\nolimits_{\mathcal{B}\subseteq\mathcal{Y}}\eta_a(\mathcal{B} \mid \mathbf{y})}\iff \eta_a(\mathcal{C} \mid \mathbf{y})=\eta_a(\mathcal{D} \mid \mathbf{y}).
\end{equation*}
To show equivalence between (i) and the model in \citet{manzini2014stochastic} assume first that $\eta_a(\cdot \mid \mathbf{y})$ is multiplicative (i.e., satisfies condition (i)). Then for every $v\in\mathcal{Y}$ define
\[
\operatorname{Q}_a(v\mid\mathbf{y})=\dfrac{\eta_a(\{y\}\mid\mathbf{y})}{1+\eta_a(\{v\}\mid\mathbf{y})}.
\] 
Since $\eta_a(\{v\}\mid\mathbf{y})>0$ (otherwise multiplicativity of $\eta_a$ would imply that sets that contain $v$ are never considered, which would contradict Assumption A1') and finite, we get that $\operatorname{Q}_a(v\mid\mathbf{y}))\in(0,1)$. Take any $A$. By multiplicativity of $\eta_a$ we get that 
\[
\eta_a(A \mid \mathbf{y})=\prod\nolimits_{y\in A}\eta_a(\{y\} \mid \mathbf{y}).
\] 
Note that for singleton $\mathcal{Y}$ we have that $\sum\nolimits_{C\subseteq\mathcal{Y}}\eta_a(C \mid \mathbf{y})=1+\eta_a(\mathcal{Y} \mid \mathbf{y})$. Then using multiplicativity we get that for $\mathcal{Y}\cup\{y^*\}$ such that $y^*\not\in \mathcal{Y}$
\begin{align*}
    \sum\nolimits_{C\subseteq\mathcal{Y}\cup\{y^*\}}\eta_a(C \mid \mathbf{y})=\sum\nolimits_{C\subseteq\mathcal{Y}}\eta_a(C \mid \mathbf{y})+\sum\nolimits_{C\subseteq\mathcal{Y}}\eta_a(C\cup\{y^*\} \mid \mathbf{y})=(1+\eta_a(\{y^*\} \mid \mathbf{y}))\sum\nolimits_{C\subseteq\mathcal{Y}}\eta_a(C \mid \mathbf{y}).
\end{align*}
Hence, by induction
\[
\sum\nolimits_{C\subseteq\mathcal{Y}}\eta_a(C \mid \mathbf{y})= \prod\nolimits_{y\in\mathcal{Y}}(1+\eta_a(\{y\} \mid \mathbf{y})).
\]
As a result,
\begin{align*}
\dfrac{\eta_a(A \mid \mathbf{y})}{\sum\nolimits_{C\subseteq\mathcal{Y}}\eta_a(C \mid \mathbf{y})}&=\prod\nolimits_{y\in A}\dfrac{\eta_a(\{y\} \mid \mathbf{y})}{1+\eta_a(\{y\} \mid \mathbf{y})}\prod\nolimits_{y'\in\mathcal{Y}\setminus A}\left(1-\dfrac{\eta_a(\{y'\} \mid \mathbf{y})}{1+\eta_a(\{y'\} \mid \mathbf{y})}\right)\\
&=\prod\nolimits_{y\in A}\operatorname{Q}_a(y \mid \mathbf{y})\prod\nolimits_{y'\in\mathcal{Y}\setminus A}(1-\operatorname{Q}_a(y' \mid \mathbf{y})).
\end{align*}
Thus, condition (i) implies the model in \citet{manzini2014stochastic}.
\par
To show the opposite assume that the probability of considering a set $\mathcal{A}$ is
\[
\prod\nolimits_{y\in A}\operatorname{Q}_a(y \mid \mathbf{y})\prod\nolimits_{y'\in\mathcal{Y}\setminus A}(1-\operatorname{Q}_a(y' \mid \mathbf{y}))
\]
for some $\operatorname{Q}_a$. For singleton sets define $\eta_a(\{y\} \mid \mathbf{y})=\dfrac{\operatorname{Q}_a(y \mid \mathbf{y})}{1-\operatorname{Q}_a(y \mid \mathbf{y})}$. For sets with cardinality bigger than one define $\eta_a(A \mid \mathbf{y})=\prod\nolimits_{y\in A}\eta_a(\{y\} \mid \mathbf{y})$. By construction, the constructed $\eta_a$ satisfies multiplicativity and it is easy to verify that 
\[ 
\prod\nolimits_{y\in A}\operatorname{Q}_a(y \mid \mathbf{y})\prod\nolimits_{y'\in\mathcal{Y}\setminus A}(1-\operatorname{Q}_a(y' \mid \mathbf{y}))=\dfrac{\eta_a(A \mid \mathbf{y})}{\sum\nolimits_{C\subseteq\mathcal{Y}}\eta_a(C \mid \mathbf{y})}.
\]

\section{Gibbs Random Field Model}\label{app: gibbs}
\noindent In this section we connect our equilibrium concept with the so-called Gibbs equilibrium. This solution concept has been extensively used in Economics to study social interactions. (See \citealp{allen1982some, blume1993statistical, blume1995statistical}, and \citealp{blume2003equilibrium}, among many others.)
The starting point of the Gibbs random field models is a set of conditional probability distributions. Each element of the set describes the probability that a given person selects each alternative as a function of the profile of choices of the other people. In our model, the set of conditional
probabilities is $\left( \operatorname{P}_{a}\right)_{a\in \mathcal{A}}$, with a generic element given by 
\begin{equation*}
\operatorname{P}_{a}\left( v \mid \mathbf{y}\right) =\operatorname{Q}_{a}\left( v \mid y_{a},\operatorname{N}_{a}^{v}\left( \mathbf{y}\right) \right) \prod\nolimits_{v'\in \mathcal{Y},v'\succ_{a}v}\left( 1-\operatorname{Q}_{a}\left( v' \mid y_{a},\operatorname{N}_{a}^{v'}\left(\mathbf{y}\right) \right) \right) \text{ for }v\in \mathcal{Y} \text{ and }\mathbf{y}\in \overline{\mathcal{Y}}^{A}.
\end{equation*}
A Gibbs equilibrium is defined as a joint distribution over the vector of choices $\mathbf{y}$, $\operatorname{P}\left( \mathbf{y}\right)$, that is able to generate $\left( \operatorname{P}_{a}\right)_{a\in \mathcal{A}}$ as its conditional distribution functions.

Gibbs equilibria typically do not exist. (In the statistical literature, a similar existence problem is referred as the issue of compatibility of conditional distributions.) The existence of Gibbs equilibria depends on a great deal of homogeneity among people. In our model, it would also require $\operatorname{Q}_{a}\left( v \mid y_{a},\operatorname{N}_{a}^{v}\left( \mathbf{y}\right)\right)$ to be invariant with respect to $y_{a}$. Condition G1 captures this restriction.

\bigskip

\noindent \textbf{(G1) }For each $a\in \mathcal{A}$, $v\in \mathcal{Y}$, and $\mathbf{y}\in \overline{\mathcal{Y}}^{A}$, $\operatorname{Q}_{a}\left( v \mid \mathbf{y}\right) \equiv \operatorname{Q}_{a}\left( v \mid \operatorname{N}_{a}^{v}\left( \mathbf{y}\right) \right)$.

\bigskip

Together with Assumption A1, this extra condition allows a simple characterization of the invariant distribution $\mu $. We describe this characterization next.

\begin{proposition}
\label{IDAN}If A1 and G1 are satisfied, then there exists a unique $\mu $.
Also, $\mu $ satisfies 
\begin{equation*}
\mu\left( \mathbf{y}\right) =\frac{1}{\sum\nolimits_{a\in \mathcal{A}} \lambda_{a}}\sum\nolimits_{a\in \mathcal{A}}\lambda_{a}\operatorname{P}_{a}\left(y_{a} \mid \mathbf{y}\right) \mu_{-a}\left( \mathbf{y}_{-a}\right) \text{ for each }\mathbf{y\in} \overline{\mathcal{Y}}^{A}.
\end{equation*}
\end{proposition}

\textbf{Proof of Proposition~\ref{IDAN}:} The characterization of $\mu$ follows as the invariant distribution satisfies the balance condition $\sum\nolimits_{\mathbf{y}'\mathbf{\in }\overline{\mathcal{Y}}^{A}}\mu \left( \mathbf{y}'\right)\operatorname{m}\left( \mathbf{y}\mid \mathbf{y}'\right) =0$ for each $\mathbf{y\in }\overline{\mathcal{Y}}^{A}$. The next steps show this claim. 
\begin{eqnarray*}
\sum\nolimits_{\mathbf{y}'\mathbf{\in }\overline{\mathcal{Y}}^{A}}\mu \left( \mathbf{y}'\right) \operatorname{m}\left( \mathbf{y}\mid \mathbf{y}'\right) &=&0 \\
\mu\left( \mathbf{y}\right) \left(-\sum\nolimits_{\mathbf{y}'\in\overline{\mathcal{Y}}^{A}\setminus \left\{ \mathbf{y}\right\} }
\operatorname{m}\left( \mathbf{y}'\mid \mathbf{y}\right) \right) +\sum\nolimits_{\mathbf{y}'\in\overline{\mathcal{Y}}^{A}\setminus \left\{ \mathbf{y}\right\} }\mu \left( \mathbf{y}'\right) \operatorname{m}\left( \mathbf{y}\mid \mathbf{y}'\right) &=&0 \\
-\mu \left( \mathbf{y}\right) \sum\nolimits_{a\in \mathcal{A}}\sum\nolimits_{y_{a}'\in \overline{\mathcal{Y}}\setminus \left\{y_{a}\right\} }\lambda_{a}\operatorname{P}_{a}\left( y_{a}' \mid \mathbf{y}\right) +\sum\nolimits_{a\in \mathcal{A}}\sum\nolimits_{y_{a}'\in \overline{\mathcal{Y}}\setminus \left\{ y_{a}\right\} }\mu\left( y_{a}',\mathbf{y}_{-a}\right) \lambda_{a}\operatorname{P}_{a}\left(y_{a} \mid y_{a}',\mathbf{y}_{-a}\right) &=&0 \\
-\mu \left( \mathbf{y}\right) \sum\nolimits_{a\in \mathcal{A}}\lambda_{a}\left( 1-\operatorname{P}_{a}\left( y_{a} \mid \mathbf{y}\right)\right) +\sum\nolimits_{a\in \mathcal{A}}\sum\nolimits_{y_{a}'\in\overline{\mathcal{Y}}\setminus \left\{ y_{a}\right\} }\mu \left(y_{a}',\mathbf{y}_{-a}\right) \lambda_{a}\operatorname{P}_{a}\left( y_{a} \mid y_{a}',\mathbf{y}_{-a}\right) &=&0 \\
\frac{1}{\sum\nolimits_{a\in \mathcal{A}}\lambda_{a}}\sum\nolimits_{a\in 
\mathcal{A}}\lambda_{a}\left\{ \sum\nolimits_{y_{a}'\in \overline{\mathcal{Y}}}\mu \left( y_{a}',\mathbf{y}_{-a}\right) \operatorname{P}_{a}\left( y_{a} \mid y_{a}',\mathbf{y}_{-a}\right)
\right\} &=&\mu \left( \mathbf{y}\right) \\
\frac{1}{\sum\nolimits_{a\in \mathcal{A}}\lambda_{a}}\sum\nolimits_{a\in\mathcal{A}}\lambda_{a}\operatorname{P}_{a}\left( y_{a} \mid \mathbf{y}\right) \mu_{-a}\left( \mathbf{y}_{-a}\right) &=&\mu \left( \mathbf{y}\right) .
\end{eqnarray*}
In moving from the fifth line to the sixth one we used the fact that, in our model, $\operatorname{P}_{a}\left( y_{a} \mid y_{a}',\mathbf{y}_{-a}\right) =\operatorname{P}_{a}\left( y_{a} \mid \mathbf{y}_{-a}\right)$ for any $y_{a}'\mathbf{\in }\overline{\mathcal{Y}}^{A}$.{\normalsize \hfill }$\blacksquare$

\bigskip 

In what follows, we will normalize the lambdas to the value of 1. From Proposition~\ref{IDAN}, $\mu $ satisfies 
\begin{equation}
\mu \left( \mathbf{y}\right) =\frac{1}{A}\sum\nolimits_{a\in \mathcal{A}}%
\operatorname{P}_{a}\left( y_{a} \mid \mathbf{y}\right) \mu_{-a}\left( 
\mathbf{y}_{-a}\right) \text{ for each }\mathbf{y\in }\overline{\mathcal{Y}}%
^{A}.  \label{q}
\end{equation}%
We next show that if $\left( \operatorname{P}_{a}\right)_{a\in A}$ is a set of compatible conditional distributions, then $\mu $ = P solves (\ref{q}). If we let $\mu $ = P, then right hand side of (\ref{q}) is 
\begin{equation*}
\frac{1}{A}\sum\nolimits_{a\in \mathcal{A}}\operatorname{P}_{a}\left( y_{a} \mid \mathbf{y}\right) \sum\nolimits_{v\in \overline{\mathcal{Y}}}\operatorname{P}\left( v,\mathbf{y}_{-a}\right) =\frac{1}{A}\sum\nolimits_{a\in \mathcal{A}}\operatorname{P}\left( \mathbf{y}\right) =\frac{A}{A}\operatorname{P}\left( \mathbf{y}\right) =\operatorname{P}\left( \mathbf{y}\right).
\end{equation*}
Also, the left hand side of (\ref{q}) is 
\begin{equation*}
\mu \left( \mathbf{y}\right) =\operatorname{P}\left( \mathbf{y}\right) .
\end{equation*}%
Thus $\mu \left( \mathbf{y}\right) =\operatorname{P}\left( \mathbf{y}\right)$ solves (\ref{q}) for each $\mathbf{y\in }\overline{\mathcal{Y}}^{A}$.

As we mentioned earlier, and assumed in the last few lines, the existence of Gibbs equilibrium also requires the set of conditional probabilities $\left( \operatorname{P}_{a}\right)_{a\in \mathcal{A}}$ to be compatible. We formalize this idea next.\bigskip

\noindent \textbf{Definition:} We say $\left( \operatorname{P}_{a}\right)_{a\in A}$ is a set of compatible conditional distributions if there exists a joint distribution $\operatorname{P}:\overline{\mathcal{Y}}^{A}\rightarrow \left[ 0,1\right]$, with $\sum\nolimits_{\mathbf{y\in }\overline{\mathcal{Y}}^{A}}\operatorname{P}\left(\mathbf{y}\right) =1$, such that 
\begin{equation*}
\operatorname{P}_{a}\left( y_{a} \mid \mathbf{y}\right) =\operatorname{P}\left(\mathbf{y}\right) /\sum\nolimits_{y_{a}\in \overline{\mathcal{Y}}}\operatorname{P}\left( \mathbf{y}\right) \text{ for each }\mathbf{y\in }\overline{\mathcal{Y}}^{A}.
\end{equation*}

The technical conditions required for a set of conditional distributions to be compatible are discussed in \citet{kaiser2000construction}. Their analysis implies that compatibility demands strong symmetric restrictions. In particular, in the two people, two actions case, \citet{arnold1989compatible} show that compatibility holds if and only if the next equality is satisfied 
\begin{equation*}
\frac{1-\operatorname{Q}_{1}\left( 1 \mid 0\right) }{\operatorname{Q}_{1}\left( 1 
 \mid 0\right) }\frac{\operatorname{Q}_{1}\left( 1 \mid 1\right) }{1-\operatorname{Q}_{1}\left( 1 \mid 1\right) }=\frac{1-\operatorname{Q}_{2}\left( 1 \mid 0\right) }{\operatorname{Q}_{2}\left( 1 \mid 0\right) }\frac{\operatorname{Q}_{2}\left( 1 \mid 1\right) }{1-\operatorname{Q}_{2}\left( 1 \mid 1\right) }.
\end{equation*}
The last result states that, under specific conditions, the Gibbs equilibrium coincides with $\mu$. A similar connection is discussed in \citet{blume2003equilibrium}. The next proposition puts together all our previous ideas.

\begin{proposition}
\label{ID4}Assume A1 and G1 hold. If $\left( \operatorname{P}_{a}\right)_{a\in A}$
is a set of compatible conditional distributions, then $\operatorname{P}_{a}\left( y_{a} 
 \mid \mathbf{y}\right) =\mu \left( \mathbf{y}\right) /\mu
_{-a}\left( \mathbf{y}_{-a}\right)$ for each $\mathbf{y\in }\overline{ 
\mathcal{Y}}^{A}$.
\end{proposition}

We next use our first example to illustrate this result.
\bigskip

\noindent \textbf{Example 1 (continued).} The conditional choice probabilities in Example 1 satisfy the compatibility requirements. Thus, there exists a joint distribution on $y_{1}$ and $y_{2}$ that can generate them as its conditional distribution functions. In this simple case, the invariant distribution $\mu $ of the dynamic revision process coincides with the Gibbs equilibrium of the model. To see this notice that from our previous results we get
\begin{eqnarray*}
\mu \left( o\right) &=&\mu \left( o,o\right) +\mu \left( o,1\right) =\mu\left( o,o\right) +\mu\left( 1,o\right) =\frac{1-\operatorname{Q}\left( 1 \mid 1\right) }{1-\operatorname{Q}\left( 1 \mid 1\right) +\operatorname{Q}\left( 1 
 \mid 0\right) }, \\
\mu \left( 1\right) &=&\mu \left( 1,1\right) +\mu \left( o,1\right) =\mu
\left( 1,1\right) +\mu \left( 1,o\right) =\frac{\operatorname{Q}\left( 1 \mid 0\right) }{1-\operatorname{Q}\left( 1 \mid 1\right) +\operatorname{Q}\left( 1 \mid 0\right) }.
\end{eqnarray*}
(Here again, we write $Q$ instead of $Q_a$ due to the symmetry across people.) Thus, the conditional distributions are 
\begin{align*}
\mu\left( 1,o\right)/\mu\left( o\right) &=\operatorname{Q}\left( 1 \mid 0\right) \text{ and } \mu\left( o,o\right)/\mu\left( o\right) =1-\operatorname{Q}\left( 1 \mid 0\right) \\
\mu\left( 1,1\right)/\mu\left( 1\right) &=\operatorname{Q}\left( 1 \mid 1\right) \text{ and } \mu\left( o,1\right)/\mu\left( 1\right) =1-\operatorname{Q}\left( 1 \mid 1\right).
\end{align*}
It follows that the conditional choice probabilities generated by equilibrium behavior coincide with the conditional choice probabilities in the model.{\normalsize \hfill }$\blacksquare $

\section{Omitted Details of Monte Carlo Simulations}\label{app: simul}

\noindent This appendix describes how we generated the data for our simulations.
Let $\lambda =\sum\nolimits_{a\in \mathcal{A}}\lambda_{a}$. We generate the data according to an iterative procedure for a fixed time period $\mathcal{T}$. The $k$-th iteration of the procedure is as follows:

\begin{itemize}
\item[(i)] Given $\mathbf{y}_{k-1}$ set $\mathbf{y}_{k}=\mathbf{y}_{k-1}$;

\item[(ii)] Generate a draw from the exponential distribution with mean $1/\lambda $ and call it $x_{k}$;

\item[(iii)] Randomly sample an agent from the set $\mathcal{A}$, such that the probability that $a$ is picked is $\lambda_{a}/\lambda$;

\item[(iv)] Given the agent selected in the previous step and the current choice configuration $\mathbf{y}_{k}$ construct a consideration set using $\operatorname{Q}_{a}$;

\item[(v)] If the consideration set is empty, then set $y_{a,k}=0$. Otherwise pick the best alternative according to the preference order of agent $a$ from the consideration set and assign it to $y_{a,k}$.
\end{itemize}

Given the initial configuration of choices $\mathbf{y}_{0}$ we applied the above algorithm till we reached $\sum\nolimits_{k}x_{k}>\mathcal{T}$ (On average the
length of the sequence is $\lambda \mathcal{T}$). Define $z_{k}=\sum\nolimits_{l\leq k}x_{l}$. The continuous time data is $\{(y_{k},z_{k})\}$. The discrete time data is obtained from the continuous time data by splitting the interval $[0,\mathcal{T}]$ into $T=[\mathcal{T}/\Delta ]$ intervals and recording the configuration of the network at every time period $t=i\Delta $, $i=0,1,\dots,[\mathcal{T}/\Delta ]$.

\section{Additional Estimation Results}\label{app: robustness}

\subsection{Different Parts of the Sample}\label{app: different parts}
\noindent Given that the Resistance is a dynamic party game, in order to analyse the behavior of consideration probabilities and preferences over time we estimated Models I(a) and I(b) presented in Section~\ref{sec: empitrical application} using two different subsamples. The first subsample consists of the observations coming from the middle half (i.e., we trowed away the first and the last quarter of observations). The second one contains observations from the second half. 
\par
The estimated consideration set probabilities for the first subsample are presented in Figures~\ref{fig: Q short,s1}-\ref{fig: Q long,s2}. The estimated preference orders are presented in Tables~\ref{table: pref sh,s1}-\ref{table: pref md,s1}.
\begin{figure}
\centering
\begin{subfigure}{.5\textwidth}
  \centering
  \begin{tikzpicture}
    \begin{axis}[
    xmin=-0.1, xmax=4.1,
    ymin=-0.1, ymax=1.1,
    axis lines = left,
    xlabel = Number of friends,
    ylabel = Consideration probability,
    legend entries={$\operatorname{Q}$},
    legend style={at={(0.1,0.8)},anchor=west},
    ]
    \addplot[smooth,mark=*] coordinates {(0,0.031) (1,0.213) (2,0.361) (3,0.415) (4,0.52)} ;
    \end{axis}
    \end{tikzpicture}
    \caption{Experiment 1}
    \label{fig: Q short1,s1}
\end{subfigure}%
\begin{subfigure}{.5\textwidth}
  \centering
  \begin{tikzpicture}
    \begin{axis}[
    xmin=-0.1, xmax=4.1,
    ymin=-0.1, ymax=1.1,
    axis lines = left,
    xlabel = Number of friends,
    ylabel = Consideration probability,
    legend entries={$\operatorname{Q}$},
    legend style={at={(0.1,0.8)},anchor=west},
    ]
    \addplot[smooth,mark=*] coordinates {(0,0.251) (1,0.352) (2,0.363) (3,0.408) (4,0.516)} ;
    \end{axis}
    \end{tikzpicture}
  \caption{Experiment 2}
  \label{fig: Q short2,s1}
\end{subfigure}
\caption{Consideration probability. Model Ia. Middle half of the samples.}
\label{fig: Q short,s1}
\end{figure}

\begin{figure}
\centering
\begin{subfigure}{.5\textwidth}
  \centering
  \begin{tikzpicture}
    \begin{axis}[
    xmin=-0.1, xmax=4.1,
    ymin=-0.1, ymax=1.1,
    axis lines = left,
    xlabel = Number of friends,
    ylabel = Consideration probability,
    legend entries={$\operatorname{Q}_1$, $\operatorname{Q}_2$, $\operatorname{Q}_3$, $\operatorname{Q}_4$, $\operatorname{Q}_5$},
    legend style={at={(0.1,0.8)},anchor=west},
    ]
    \addplot +[smooth] coordinates {(0,0.0) (1,0.144) (2,0.306) (3,0.311) (4,0.572)};
    \addplot +[smooth] coordinates {(0,0.04) (1,0.188) (2,0.252) (3,0.567) (4,0.794)};
    \addplot +[smooth] coordinates {(0,0.008) (1,0.214) (2,0.499) (3,0.514) (4,0.69)};
    \addplot +[smooth] coordinates {(0,0.0) (1,0.428) (2,0.699) (3,0.587) (4,0.643)};
    \addplot +[smooth] coordinates {(0,0.445) (1,0.352) (2,0.321) (3,0.335) (4,0.187)};
    \end{axis}
    \end{tikzpicture}
    \caption{Experiment 1}
    \label{fig: Q long1,s1}
\end{subfigure}%
 
\begin{subfigure}{.5\textwidth}
  \centering
  \begin{tikzpicture}
    \begin{axis}[
    xmin=-0.1, xmax=4.1,
    ymin=-0.1, ymax=1.1,
    axis lines = left,
    xlabel = Number of friends,
    ylabel = Consideration probability,
    legend entries={$\operatorname{Q}_1$, $\operatorname{Q}_2$, $\operatorname{Q}_3$, $\operatorname{Q}_4$, $\operatorname{Q}_5$},
    legend style={at={(0.1,0.8)},anchor=west},
    ]
    \addplot +[smooth] coordinates {(0,0.347) (1,0.425) (2,0.389) (3,0.412) (4,0.435)};
    \addplot +[smooth] coordinates {(0,0.482) (1,0.447) (2,0.453) (3,0.412) (4,0.562)};
    \addplot +[smooth] coordinates {(0,0.098) (1,0.206) (2,0.231) (3,0.307) (4,0.495)};
    \addplot +[smooth] coordinates {(0,0.222) (1,0.308) (2,0.322) (3,0.451) (4,0.639)};
    \addplot +[smooth] coordinates {(0,0.293) (1,0.364) (2,0.403) (3,0.44) (4,0.534)};
    \end{axis}
    \end{tikzpicture}
    \caption{Experiment 2}
    \label{fig: Q long2,s1}
\end{subfigure}
\caption{Consideration probabilities. Model I(b). Middle half of the samples.}
\label{fig: Q long,s1}
\end{figure}

\begin{figure}
\centering
\begin{subfigure}{.5\textwidth}
  \centering
  \begin{tikzpicture}
    \begin{axis}[
    xmin=-0.1, xmax=4.1,
    ymin=-0.1, ymax=1.1,
    axis lines = left,
    xlabel = Number of friends,
    ylabel = Consideration probability,
    legend entries={$\operatorname{Q}$},
    legend style={at={(0.1,0.8)},anchor=west},
    ]
    \addplot[smooth,mark=*] coordinates {(0,0.039) (1,0.213) (2,0.373) (3,0.421) (4,0.506)} ;
    \end{axis}
    \end{tikzpicture}
    \caption{Experiment 1}
    \label{fig: Q short1,s2}
\end{subfigure}%
\begin{subfigure}{.5\textwidth}
  \centering
  \begin{tikzpicture}
    \begin{axis}[
    xmin=-0.1, xmax=4.1,
    ymin=-0.1, ymax=1.1,
    axis lines = left,
    xlabel = Number of friends,
    ylabel = Consideration probability,
    legend entries={$\operatorname{Q}$},
    legend style={at={(0.1,0.8)},anchor=west},
    ]
    \addplot[smooth,mark=*] coordinates {(0,0.271) (1,0.343) (2,0.358) (3,0.403) (4,0.513)} ;
    \end{axis}
    \end{tikzpicture}
  \caption{Experiment 2}
  \label{fig: Q short2,s2}
\end{subfigure}
\caption{Consideration probability. Model Ia. Second half of the samples. }
\label{fig: Q short,s2}
\end{figure}

\begin{figure}
\centering
\begin{subfigure}{.5\textwidth}
  \centering
  \begin{tikzpicture}
    \begin{axis}[
    xmin=-0.1, xmax=4.1,
    ymin=-0.1, ymax=1.1,
    axis lines = left,
    xlabel = Number of friends,
    ylabel = Consideration probability,
    legend entries={$\operatorname{Q}_1$, $\operatorname{Q}_2$, $\operatorname{Q}_3$, $\operatorname{Q}_4$, $\operatorname{Q}_5$},
    legend style={at={(0.1,0.8)},anchor=west},
    ]
    \addplot +[smooth] coordinates {(0,0.0) (1,0.155) (2,0.308) (3,0.268) (4,0.524)};
    \addplot +[smooth] coordinates {(0,0.028) (1,0.155) (2,0.28) (3,0.59) (4,0.757)};
    \addplot +[smooth] coordinates {(0,0.011) (1,0.153) (2,0.465) (3,0.521) (4,0.753)};
    \addplot +[smooth] coordinates {(0,0.151) (1,0.464) (2,0.657) (3,0.557) (4,0.565)};
    \addplot +[smooth] coordinates {(0,0.386) (1,0.386) (2,0.366) (3,0.307) (4,0.192)};
    \end{axis}
    \end{tikzpicture}
    \caption{Experiment 1}
    \label{fig: Q long1,s2}
\end{subfigure}%
\begin{subfigure}{.5\textwidth}
  \centering
  \begin{tikzpicture}
    \begin{axis}[
    xmin=-0.1, xmax=4.1,
    ymin=-0.1, ymax=1.1,
    axis lines = left,
    xlabel = Number of friends,
    ylabel = Consideration probability,
    legend entries={$\operatorname{Q}_1$, $\operatorname{Q}_2$, $\operatorname{Q}_3$, $\operatorname{Q}_4$, $\operatorname{Q}_5$},
    legend style={at={(0.1,0.8)},anchor=west},
    ]
    \addplot +[smooth] coordinates {(0,0.419) (1,0.453) (2,0.391) (3,0.388) (4,0.43)};
    \addplot +[smooth] coordinates {(0,0.465) (1,0.448) (2,0.417) (3,0.436) (4,0.656)};
    \addplot +[smooth] coordinates {(0,0.133) (1,0.175) (2,0.225) (3,0.268) (4,0.438)};
    \addplot +[smooth] coordinates {(0,0.187) (1,0.258) (2,0.316) (3,0.44) (4,0.623)};
    \addplot +[smooth] coordinates {(0,0.395) (1,0.389) (2,0.445) (3,0.48) (4,0.542)};
    \end{axis}
    \end{tikzpicture}
    \caption{Experiment 2}
    \label{fig: Q long2,s2}
\end{subfigure}
\caption{Consideration probabilities. Model I(b). Second half of the samples.}
\label{fig: Q long,s2}
\end{figure}

\begin{table}[h]
\centering
\begin{threeparttable}
\centering
\caption{Preferences. Model I(a).}\label{table: pref sh,s1}
\begin{tabular}{ccccccc}
\hline
\hline
&                   & Player 1 & Player 2 & Player 3 & Player 4 & Player 5\\ 
\hline
Middle half&Experiment 1       & L,T,R     & L,T,R     & L,T,R     & L,T,R     & L,T,R    \\
&Experiment 2       & L,T,R    & L,T,R    & L,T,R    & L,T,R    & L,T,R   \\
2nd half &Experiment 1       & L,T,R     & L,T,R     & L,T,R     & L,T,R     & L,T,R    \\
&Experiment 2       & L,T,R     & L,T,R     & L,T,R     & L,T,R     & L,T,R    \\
\hline
\end{tabular}
\vspace{1ex}
\end{threeparttable}
\end{table}

\begin{table}[h]
\centering
\begin{threeparttable}
\centering
\caption{Preferences. Model I(b).}\label{table: pref md,s1}
\begin{tabular}{ccccccc}
\hline
\hline
                  & Player 1 & Player 2 & Player 3 & Player 4 & Player 5\\ 
\hline
Middle half&Experiment 1       & L,T,R     & L,T,R     & L,T,R     & L,T,R     & L,T,R    \\
&Experiment 2       & L,T,R    & L,T,R    & L,T,R    & L,T,R    & L,T,R   \\
2nd half &Experiment 1       & L,T,R     & L,T,R     & L,T,R     & L,T,R     & L,T,R    \\
&Experiment 2       & L,T,R     & L,T,R     & L,T,R     & L,T,R     & L,T,R    \\
\hline
\end{tabular}
\vspace{1ex}
\end{threeparttable}
\end{table}

\subsection{Heterogeneity in Consideration Probabilities Across Choices}\label{app: diff spec}
In this section we present results of estimation of the following specifications of the model:
\begin{enumerate}
    \item Model Ic: $\operatorname{Q}
    _{a}\left( v \mid y_{a},\operatorname{N}_{a}^{v}\left( \mathbf{y}\right)
    \right)=\operatorname{Q}\left( v \mid \operatorname{N}_{a}^{v}\left( \mathbf{y}\right)
    \right)$ for all $a,v,\mathbf{y}$;
    \item Model IIa: $\operatorname{Q}
    _{a}\left( v \mid y_{a},\operatorname{N}_{a}^{v}\left( \mathbf{y}\right)
    \right)=\operatorname{Q}\left(\text{State},\operatorname{N}_{a}^{v}\left( \mathbf{y}\right)\right)$ for all $a,v,\mathbf{y}$. State equals to $1$ if $v=y_a$, and $0$ otherwise.
    \item Model IIb: $\operatorname{Q}
    _{a}\left( v \mid y_{a},\operatorname{N}_{a}^{v}\left( \mathbf{y}\right)
    \right)=\operatorname{Q}_a\left(\text{State},\operatorname{N}_{a}^{v}\left( \mathbf{y}\right)\right)$ for all $a,v,\mathbf{y}$. State equals to $1$ if $v=y_a$, and $0$ otherwise.
\end{enumerate}

The estimated consideration probabilities are presented in Figures~\ref{fig: Q mIc}-~\ref{fig: Q mIIb2} and Tables~\ref{table: pref mIc}-~\ref{table: pref mIIb}.

\begin{figure}
\centering
\begin{subfigure}{.5\textwidth}
  \centering
  \begin{tikzpicture}
    \begin{axis}[
    xmin=-0.1, xmax=4.1,
    ymin=-0.1, ymax=1.1,
    axis lines = left,
    xlabel = Number of friends,
    ylabel = Consideration probability,
    legend entries={$\operatorname{Q}(t)$, $\operatorname{Q}(l)$, $\operatorname{Q}(r)$},
    legend style={at={(0.1,0.8)},anchor=west},
    ]
    \addplot +[smooth] coordinates {(0,0.016) (1,0.028) (2,0.049) (3,0.031) (4,0.056)};
    \addplot +[smooth] coordinates {(0,0.341) (1,0.365) (2,0.463) (3,0.576) (4,0.757)};
    \addplot +[smooth] coordinates {(0,1) (1,0.786) (2,1) (3,1) (4,1)};
    \end{axis}
    \end{tikzpicture}
    \caption{Experiment 1}
    \label{fig: Q mIce1}
\end{subfigure}%
\begin{subfigure}{.5\textwidth}
  \centering
  \begin{tikzpicture}
    \begin{axis}[
    xmin=-0.1, xmax=4.1,
    ymin=-0.1, ymax=1.1,
    axis lines = left,
    xlabel = Number of friends,
    ylabel = Consideration probability,
    legend entries={$\operatorname{Q}(t)$, $\operatorname{Q}(l)$, $\operatorname{Q}(r)$},
    legend style={at={(0.1,0.8)},anchor=west},
    ]
    \addplot +[smooth] coordinates {(0,0.26) (1,0.213) (2,0.218) (3,0.183) (4,0.287)};
    \addplot +[smooth] coordinates {(0,0.19) (1,0.389) (2,0.421) (3,0.518) (4,0.671)};
    \addplot +[smooth] coordinates {(0,1) (1,1) (2,1) (3,1) (4,1)};
    \end{axis}
    \end{tikzpicture}
    \caption{Experiment 2}
    \label{fig: Q mIce2}
\end{subfigure}
\caption{Consideration probabilities. Model Ic.}
\label{fig: Q mIc}
\end{figure}

\begin{table}[h]
\centering
\begin{threeparttable}
\centering
\caption{Preferences. Model Ic}\label{table: pref mIc}
\begin{tabular}{cccccc}
\hline
\hline
                   & Player 1 & Player 2 & Player 3 & Player 4 & Player 5\\ 
\hline
Experiment 1       & T,L,R     & T,L,R     & L,T,R     & T,L,R     & L,T,R    \\
Experiment 2       & T,L,R     & T,L,R     & L,T,R     & T,L,R     & T,L,R    \\
\hline
\end{tabular}
\vspace{1ex}
\end{threeparttable}
\end{table}

\begin{figure}
\centering
\begin{subfigure}{.5\textwidth}
  \centering
  \begin{tikzpicture}
    \begin{axis}[
    xmin=-0.1, xmax=4.1,
    ymin=-0.1, ymax=1.1,
    axis lines = left,
    xlabel = Number of friends,
    ylabel = Consideration probability,
    legend entries={$\text{State}=1$, $\text{State}=0$},
    legend style={at={(0.1,0.8)},anchor=west},
    ]
    \addplot +[smooth] coordinates {(0,0.0) (1,0.0) (2,0.0) (3,0.069) (4,0.441)};
    \addplot +[smooth] coordinates {(0,0.027) (1,0.21) (2,0.354) (3,0.418) (4,0.453)};
    \end{axis}
    \end{tikzpicture}
    \caption{Experiment 1}
    \label{fig: Q mIIae1}
\end{subfigure}%
\begin{subfigure}{.5\textwidth}
  \centering
  \begin{tikzpicture}
    \begin{axis}[
    xmin=-0.1, xmax=4.1,
    ymin=-0.1, ymax=1.1,
    axis lines = left,
    xlabel = Number of friends,
    ylabel = Consideration probability,
    legend entries={$\text{State}=1$, $\text{State}=0$},
    legend style={at={(0.1,0.8)},anchor=west},
    ]
    \addplot +[smooth] coordinates {(0,0.0) (1,0.0) (2,0.0) (3,0.0) (4,0.0)};
    \addplot +[smooth] coordinates {(0,0.308) (1,0.338) (2,0.381) (3,0.448) (4,0.664)};
    \end{axis}
    \end{tikzpicture}
    \caption{Experiment 2}
    \label{fig: Q mIIae2}
\end{subfigure}
\caption{Consideration probabilities. Model IIa.}
\label{fig: Q mIIa}
\end{figure}

\begin{table}[h]
\centering
\begin{threeparttable}
\centering
\caption{Preferences. Model IIa}\label{table: pref mIII}
\begin{tabular}{cccccc}
\hline
\hline
                   & Player 1 & Player 2 & Player 3 & Player 4 & Player 5\\ 
\hline
Experiment 1       & L,T,R     & L,T,R     & L,T,R     & L,T,R     & L,T,R    \\
Experiment 2       & L,T,R     & L,T,R     & L,T,R     & L,T,R     & L,T,R    \\
\hline
\end{tabular}
\vspace{1ex}
\end{threeparttable}
\end{table}

\begin{figure}
\centering
\begin{subfigure}{.5\textwidth}
  \centering
  \begin{tikzpicture}
    \begin{axis}[
    xmin=-0.1, xmax=4.1,
    ymin=-0.1, ymax=1.1,
    axis lines = left,
    xlabel = Number of friends,
    ylabel = Consideration probability,
    legend entries={$\operatorname{Q}_1$, $\operatorname{Q}_2$, $\operatorname{Q}_3$, $\operatorname{Q}_4$,$\operatorname{Q}_5$},
    legend style={at={(0.1,0.8)},anchor=west},
    ]
    \addplot +[smooth] coordinates {(0,0.0) (1,0.0) (2,0.0) (3,0.0) (4,0.0)};
    \addplot +[smooth] coordinates {(0,0.0) (1,0.0) (2,0.11) (3,0.469) (4,0.226)};
    \addplot +[smooth] coordinates {(0,0.0) (1,0.0) (2,0.0) (3,0.0) (4,0.52)};
    \addplot +[smooth] coordinates {(0,1.0) (1,0.644) (2,0.554) (3,0.679) (4,0.775)};
    \addplot +[smooth] coordinates {(0,0.0) (1,0.0) (2,0.0) (3,0.516) (4,0.457)};
    \end{axis}
    \end{tikzpicture}
    \caption{State$=1$}
    \label{fig: Q mIIbe1s1}
\end{subfigure}%
\begin{subfigure}{.5\textwidth}
  \centering
  \begin{tikzpicture}
    \begin{axis}[
    xmin=-0.1, xmax=4.1,
    ymin=-0.1, ymax=1.1,
    axis lines = left,
    xlabel = Number of friends,
    ylabel = Consideration probability,
    legend entries={$\operatorname{Q}_1$, $\operatorname{Q}_2$, $\operatorname{Q}_3$, $\operatorname{Q}_4$,$\operatorname{Q}_5$},
    legend style={at={(0.1,0.8)},anchor=west},
    ]
    \addplot +[smooth] coordinates {(0,0.0) (1,0.154) (2,0.384) (3,0.319) (4,0.457)};
    \addplot +[smooth] coordinates {(0,0.059) (1,0.218) (2,0.226) (3,0.585) (4,0.879)};
    \addplot +[smooth] coordinates {(0,0.003) (1,0.182) (2,0.511) (3,0.665) (4,0.789)};
    \addplot +[smooth] coordinates {(0,0.102) (1,0.362) (2,0.629) (3,0.498) (4,0.392)};
    \addplot +[smooth] coordinates {(0,0.368) (1,0.432) (2,0.351) (3,0.305) (4,0.195)};
    \end{axis}
    \end{tikzpicture}
    \caption{State$=0$}
    \label{fig: Q mIIbe1s2}
\end{subfigure}
\caption{Consideration probabilities. Model IIb. Experiment 1.}
\label{fig: Q mIIb1}
\end{figure}

\begin{figure}
\centering
\begin{subfigure}{.5\textwidth}
  \centering
  \begin{tikzpicture}
    \begin{axis}[
    xmin=-0.1, xmax=4.1,
    ymin=-0.1, ymax=1.1,
    axis lines = left,
    xlabel = Number of friends,
    ylabel = Consideration probability,
    legend entries={$\operatorname{Q}_1$, $\operatorname{Q}_2$, $\operatorname{Q}_3$, $\operatorname{Q}_4$,$\operatorname{Q}_5$},
    legend style={at={(0.1,0.8)},anchor=west},
    ]
    \addplot +[smooth] coordinates {(0,0.0) (1,0.0) (2,0.0) (3,0.0) (4,0.0)};
    \addplot +[smooth] coordinates {(0,0.0) (1,0.0) (2,0.0) (3,0.0) (4,0.0)};
    \addplot +[smooth] coordinates {(0,0.0) (1,0.0) (2,0.0) (3,0.0) (4,0.0)};
    \addplot +[smooth] coordinates {(0,0.0) (1,0.0) (2,0.0) (3,0.0) (4,0.0)};
    \addplot +[smooth] coordinates {(0,0.156) (1,0.0) (2,0.138) (3,0.132) (4,0.383)};
    \end{axis}
    \end{tikzpicture}
    \caption{State$=1$}
    \label{fig: Q mIIbe2s1}
\end{subfigure}%
\begin{subfigure}{.5\textwidth}
  \centering
  \begin{tikzpicture}
    \begin{axis}[
    xmin=-0.1, xmax=4.1,
    ymin=-0.1, ymax=1.1,
    axis lines = left,
    xlabel = Number of friends,
    ylabel = Consideration probability,
    legend entries={$\operatorname{Q}_1$, $\operatorname{Q}_2$, $\operatorname{Q}_3$, $\operatorname{Q}_4$,$\operatorname{Q}_5$},
    legend style={at={(0.1,0.8)},anchor=west},
    ]
    \addplot +[smooth] coordinates {(0,0.395) (1,0.432) (2,0.417) (3,0.455) (4,0.653)};
    \addplot +[smooth] coordinates {(0,0.59) (1,0.567) (2,0.537) (3,0.52) (4,0.775)};
    \addplot +[smooth] coordinates {(0,0.166) (1,0.182) (2,0.232) (3,0.332) (4,0.517)};
    \addplot +[smooth] coordinates {(0,0.291) (1,0.297) (2,0.385) (3,0.519) (4,0.766)};
    \addplot +[smooth] coordinates {(0,0.37) (1,0.354) (2,0.394) (3,0.464) (4,0.606)};
    \end{axis}
    \end{tikzpicture}
    \caption{State$=0$}
    \label{fig: Q mIIbe2s2}
\end{subfigure}
\caption{Consideration probabilities. Model IIb. Experiment 2.}
\label{fig: Q mIIb2}
\end{figure}

\begin{table}[h]
\centering
\begin{threeparttable}
\centering
\caption{Preferences. Model IIb}\label{table: pref mIIb}
\begin{tabular}{cccccc}
\hline
\hline
                   & Player 1 & Player 2 & Player 3 & Player 4 & Player 5\\ 
\hline
Experiment 1       & L,T,R     & L,T,R     & L,T,R     & L,T,R     & L,T,R    \\
Experiment 2       & L,T,R     & L,T,R     & L,T,R     & L,T,R     & L,T,R    \\
\hline
\end{tabular}
\vspace{1ex}
\end{threeparttable}
\end{table}

\end{document}